\pdfoutput=1
\newcommand*{\ATLASLATEXPATH}{latex/}
\documentclass[cernpreprint, texlive=2016, UKenglish]{\ATLASLATEXPATH atlasdoc}

\usepackage{\ATLASLATEXPATH atlaspackage}
\usepackage{\ATLASLATEXPATH atlasbiblatex}
\usepackage{\ATLASLATEXPATH atlasphysics}
\usepackage{placeins}
\usepackage{multirow}
\usepackage{pdflscape}
\usepackage{float}
\graphicspath{{logos/}{figures/}}
\usepackage{ANA-STDM-2017-22-PAPER-defs}

\AtlasTitle{Evidence for the production of three massive vector bosons with the
ATLAS detector}
\AtlasAbstract{%
A search for the production of three massive vector bosons in proton--proton
collisions is performed using data at $\sqrt{s} = 13$ TeV recorded with the
ATLAS detector at the Large Hadron Collider in the years 2015--2017,
corresponding to an integrated luminosity of $79.8$ fb$^{-1}$. Events with two
same-sign leptons $\ell$ (electrons or muons) and at least two reconstructed
jets are selected to search for $WWW \to \ell \nu \ell \nu qq$. Events with
three leptons without any same-flavour opposite-sign lepton pairs are used to
search for $WWW \to \ell \nu \ell\nu \ell \nu$, while events with three leptons
and at least one same-flavour opposite-sign lepton pair and one or more
reconstructed jets are used to search for $WWZ \to \ell \nu qq \ell \ell$.
Finally, events with four leptons are analysed to search for $WWZ \to \ell \nu
\ell \nu \ell \ell$ and $WZZ \to qq \ell \ell \ell \ell$. Evidence for the
joint production of three massive vector bosons is observed with a significance
of 4.1 standard deviations, where the expectation is 3.1 standard deviations.}
\author{The ATLAS Collaboration}
\PreprintIdNumber{CERN-EP-2019-041}
\AtlasRefCode{STDM-2017-22} 
\AtlasJournalRef{Phys.\ Lett.\ B 798 (2019) 134913}
\AtlasDOI{10.1016/j.physletb.2019.134913}

\hypersetup{pdftitle={ATLAS document},pdfauthor={The ATLAS Collaboration}}
\begin{document}
\maketitle

\section{Introduction}

The joint production of three vector bosons is a rare process in the Standard
Model (SM).  Studies of triboson production can test the non-Abelian gauge
structure of the SM theory and any deviations from the SM prediction would
provide hints of new physics at higher energy scales~\cite{Belyaev:1998ih,
Du:2012vh, Fichet:2013ola, Giudice:2007fh}.  Triboson production has been
studied at the Large Hadron Collider (LHC) using proton--proton ($pp$)
collision data taken at $\sqrt{s}= \SI{8}{\TeV}$ for processes such as
$\gamma\gamma\gamma$~\cite{STDM-2016-06}, \Wgg~\cite{STDM-2013-05,
CMS-SMP-15-008}, $Z\gamma\gamma$~\cite{STDM-2014-01, CMS-SMP-15-008}, \WWg and
\WZg~\cite{CMS-SMP-13-009, STDM-2016-05}, and \WWW~\cite{STDM-2015-07}. 

\begin{figure}[htbp]
\centering
\includegraphics[width=0.23\figwidth]{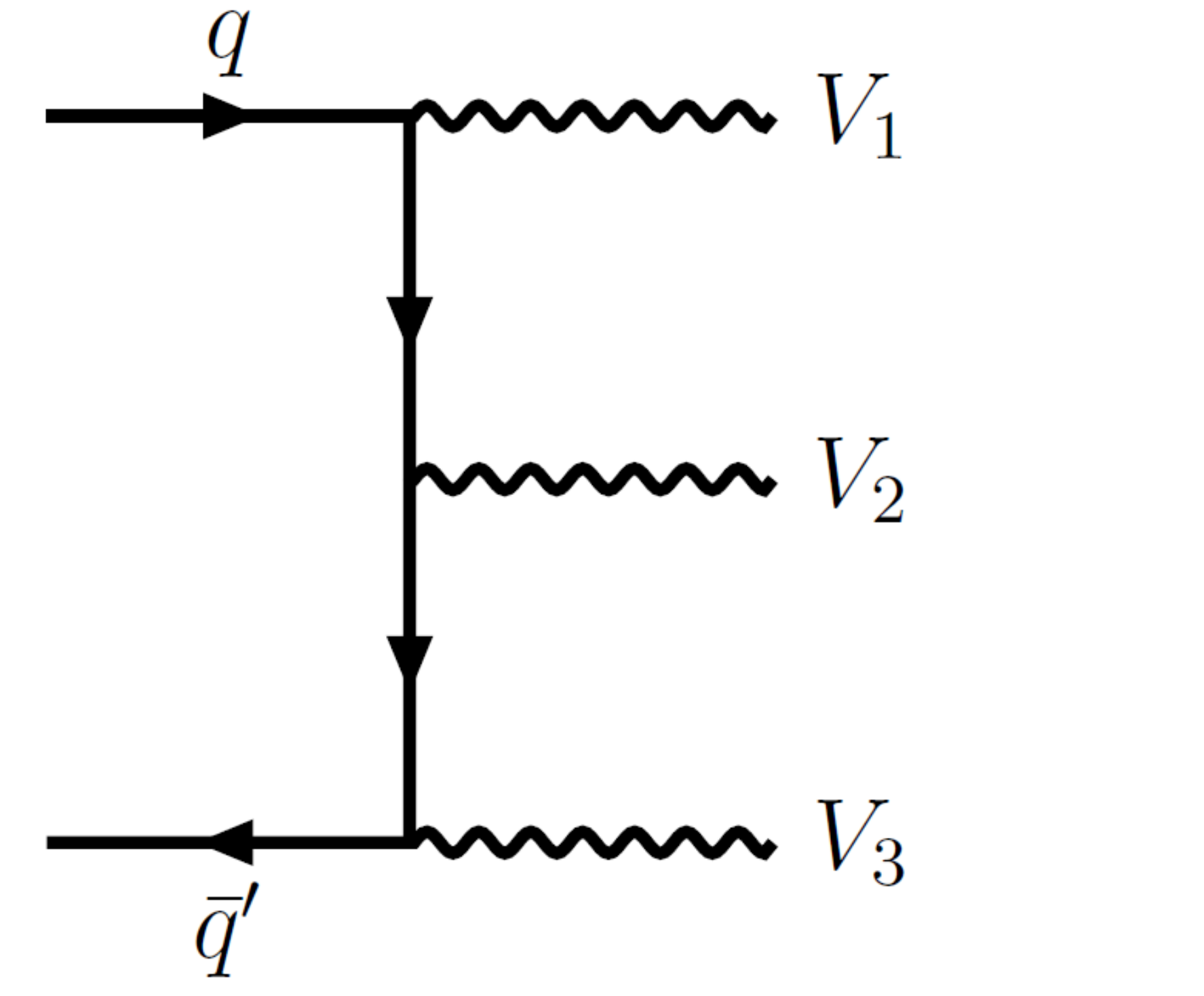} \hspace{0.05em}
\includegraphics[width=0.28\figwidth]{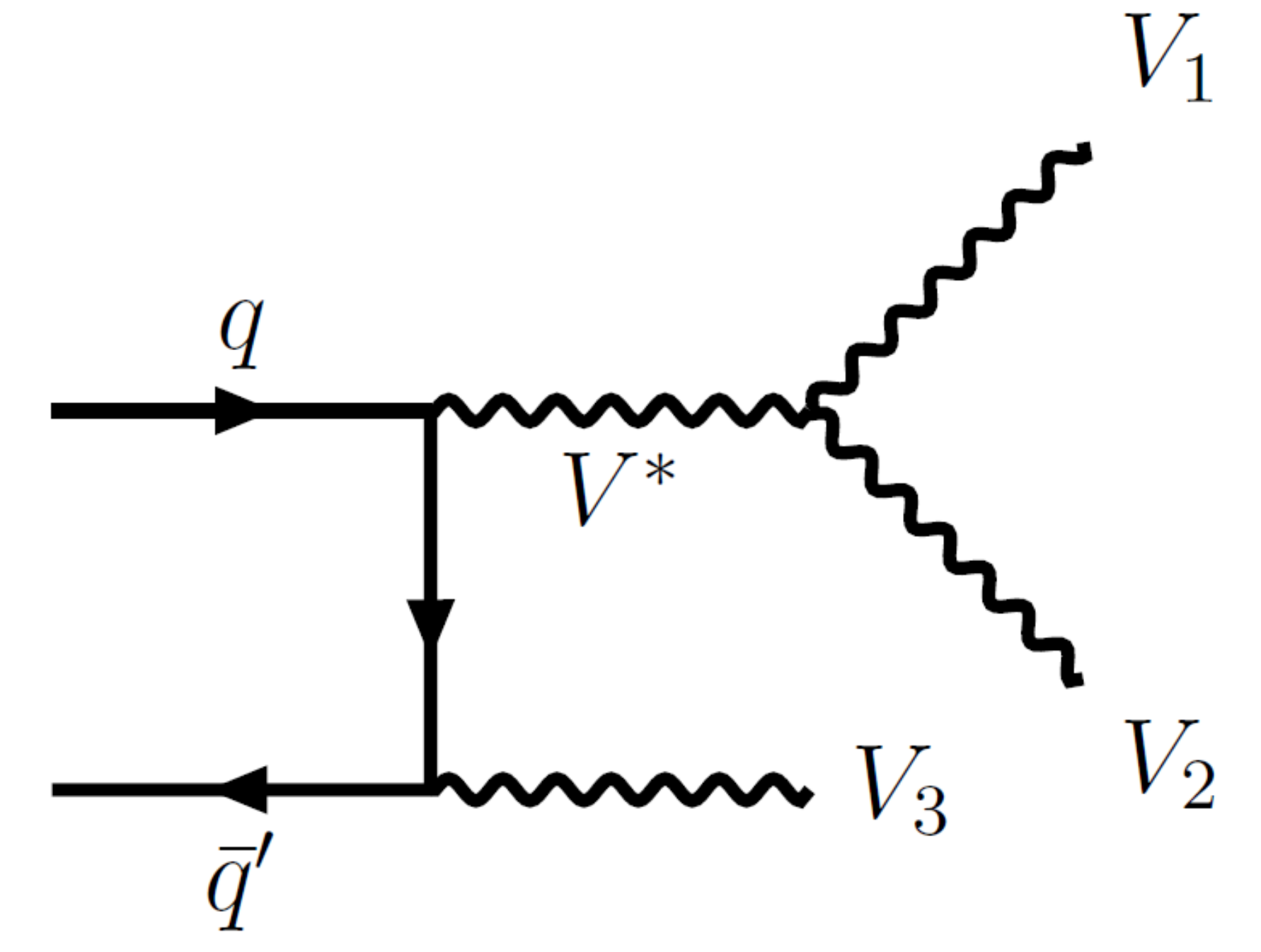} \hspace{0.5em}
\includegraphics[width=0.26\figwidth]{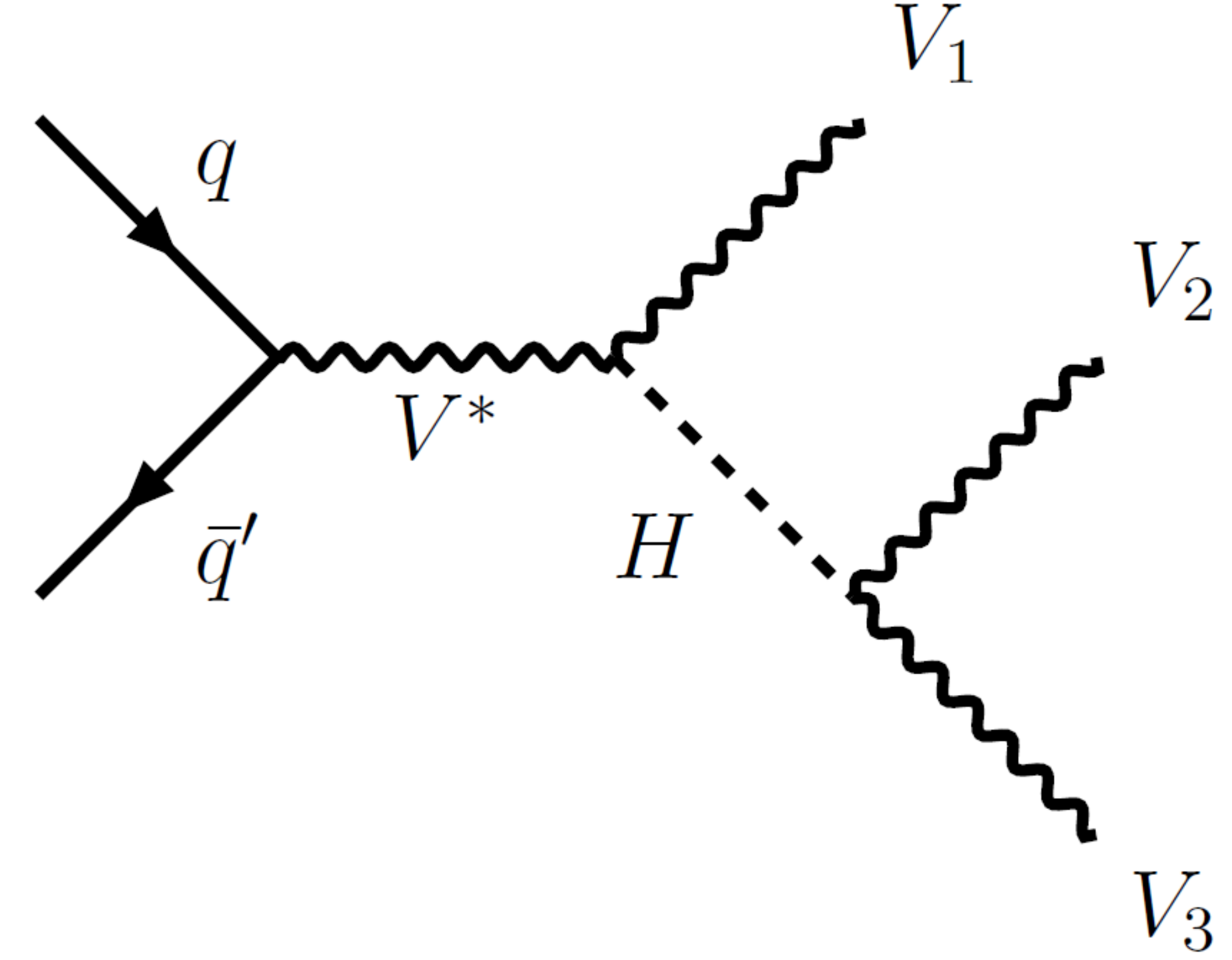} \hspace{0.5em}
\includegraphics[width=0.28\figwidth]{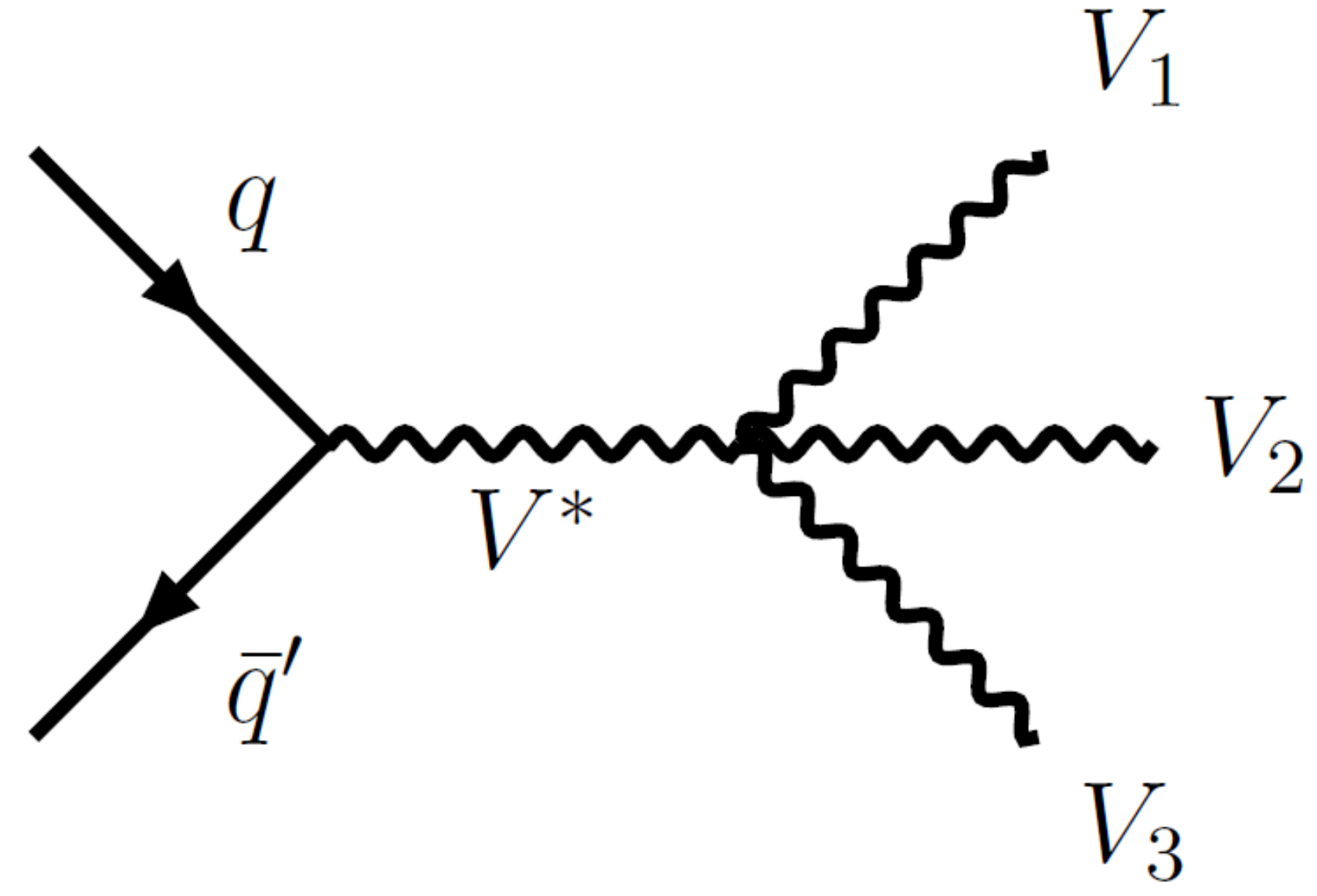} 
\caption{Representative Feynman diagrams at LO for the production of three
massive vector bosons, including diagrams sensitive to triple and quartic gauge
couplings.}
\label{fig:feyn}
\end{figure}

This letter presents the first evidence for the joint production of three
massive vector bosons in $pp$ collisions using the dataset collected with the
ATLAS detector between 2015 and 2017 at $\sqrt{s} = \SI{13}{\TeV}$.  At leading
order (LO) in quantum chromodynamics (QCD), the production of three massive
vector bosons ($VVV$, with $V=W, Z$) can proceed via the radiation of each
vector boson from a fermion, from an associated boson production with an
intermediate boson ($W$, $Z/\gamma^*$ or $H$) decaying into two vector bosons,
or from a quartic gauge coupling vertex.  Representative Feynman diagrams are
shown in Figure~\ref{fig:feyn}.

Two dedicated searches are performed, one for the \WWWs (denoted as \WWW)
process and one for the \WWZs (denoted as \WWZ) and \WZZs (denoted as \WZZ)
processes.  To search for the \WWW process, events with two same-sign leptons
with at least two jets resulting from \sswww ($\ell=e, \mu$, including $\tau
\to \ell \nu \nu$) or three leptons resulting from \lllwww are considered and
are hereafter referred to as the \lljj and \lllv channels, respectively.  To
search for the \WWZ and \WZZ (denoted as \WVZ) processes, events with three or
four leptons resulting from \lllwvz, \llllwwz, and \llllwzz are used. Selection
criteria are chosen in order to ensure there is no overlap between different
channels.  A discriminant that separates the \WWW or \WVZ signal from the background is 
defined in each channel.
The discriminants are combined using a binned
maximum-likelihood fit, which allows the signal yield and the background
normalisations to be extracted.  The combined observable is the signal strength
parameter $\mu$ defined as the ratio of the measured $WVV$ cross section to its
SM expectation, where one common ratio is assumed for $WWW$ and $WVZ$.

\section{The ATLAS detector, data and simulation samples}
\label{s:detector}

The ATLAS detector~\cite{PERF-2007-01, ATLAS-TDR-19, Abbott:2018ikt} is a
multi-purpose particle detector comprised of an inner detector (ID) surrounded
by a \SI{2}{T} superconducting solenoid, electromagnetic (EM) and hadronic
calorimeters, and a muon spectrometer (MS) with one barrel and two endcap
air-core toroids.  The ID consists of a silicon pixel detector, a silicon
microstrip detector, and a transition radiation tracker, and covers $|\eta| <
2.5$ in pseudorapidity.\footnote{ATLAS uses a right-handed coordinate system
with its origin at the nominal interaction point (IP) in the centre of the
detector and the $z$-axis along the beam pipe. The $x$-axis points from the IP
to the centre of the LHC ring, and the $y$-axis points upward. Cylindrical
coordinates $(r,\phi)$ are used in the transverse plane, $\phi$ being the
azimuthal angle around the beam pipe. The pseudorapidity is defined in terms of
the polar angle $\theta$ as $\eta=-\ln\tan(\theta/2)$. Angular distance is
measured in units of $\Delta R=\sqrt{(\Delta \eta)^2+(\Delta \phi)^2}$.} The
calorimeter system covers the pseudorapidity range $|\eta| < 4.9$.  The MS
provides muon triggering capability for $|\eta|<2.4$ and muon identification
and measurement for $|\eta| < 2.7$.  A two-level trigger
system~\cite{TRIG-2016-01}, using custom hardware followed by a software-based
trigger level, is used to reduce the event rate to an average of around
\SI{1}{kHz} for offline storage.

The data used were collected between 2015 and 2017 in $pp$ collisions at
$\sqrt{s} = \SI{13}{\TeV}$.  Only events recorded with a fully operational
detector and stable beams are included.  Candidate events are selected by
single isolated-lepton ($e$ or $\mu$) triggers with transverse momentum
thresholds varying from $\pt = \SI{20}{\GeV}$ to $\SI{26}{\GeV}$ (depending on
the lepton flavour and run period) or single-lepton triggers with thresholds of
$\pt = \SI{50}{\GeV}$ for muons and $\pt = \SI{60}{\GeV}$ for electrons.  
Due to the presence of two, three or four leptons in the final state, 
these single-lepton triggers are fully efficient for the triboson signals 
in the signal regions defined in Sections~\ref{s:www} and~\ref{s:wvz}.
The resulting total integrated luminosity is \lumi. 

Signal and background processes were simulated with several Monte Carlo (MC)
event generators, while the ATLAS detector response was
modelled~\cite{SOFT-2010-01} with {\textsc{Geant4}}~\cite{geant}.  The effect
of multiple $pp$ interactions in the same and neighbouring bunch crossings
(pile-up) was included by overlaying minimum-bias events simulated with
\PYTHIAV{8.186}~\cite{Sjostrand:2007gs} interfaced to
\textsc{EvtGen}~1.2.0~\cite{evtgen}, referred to as \PYTHIAV{8.1} in the
following, and using the \tune{A3}~\cite{ATL-PHYS-PUB-2016-017} set of tuned MC
parameters, on each generated event in all samples.  Triboson signal events~\cite{ATL-PHYS-PUB-2017-005}
were generated using \SHERPAV{2.2.2}~\cite{sherpa, Gleisberg:2008fv,
Cascioli:2011va} with the \pdf{NNPDF3.0NNLO}~\cite{nnpdf} parton distribution
function (PDF) set, where all three bosons are on-mass-shell, using a
factorised approach~\cite{ATL-PHYS-PUB-2016-002}.  Events with an off-mass-shell
boson through $WH \rightarrow WVV^*$ and $ZH \rightarrow ZVV^*$ were generated
using \POWHEGBOXV{2}~\cite{powheg, Nason:2004rx, Frixione:2007vw,
Cullen:2011ac, Hamilton:2012np, Luisoni:2013kna} interfaced to \PYTHIAV{8.1}
for the $WWW$ analysis, while for the \WVZ analysis only \PYTHIAV{8.1} was
used.  The generator was interfaced to the \pdf{CT10}~\cite{Pumplin:2002vw}
(\pdf{NNPDF2.3LO}~\cite{Ball:2012cx}) PDF and the
\tune{AZNLO}~\cite{STDM-2012-23} (\tune{A14}~\cite{ATL-PHYS-PUB-2014-021}) set
of tuned MC parameters for the \WWW (\WVZ) analysis.  Both on-mass-shell and
off-mass-shell processes were generated at next-to-leading order (NLO) QCD
accuracy~\cite{Lazopoulos:2007ix, Binoth:2008kt, Hankele:2007sb,
Campanario:2008yg} and are included in the signal definition. The expected
cross sections for $WWW$ and $WWZ$ production are \SI{0.50}{pb} and
\SI{0.29}{pb}, respectively, with an uncertainty of $\sim\!10\,\%$, evaluated
by varying parameters in the simulation related to the 
renormalisation and factorisation scales, parton shower and PDF sets.

Diboson ($WW$, $WZ$, $ZZ$)~\cite{ATL-PHYS-PUB-2016-002},
$W/Z+\gamma$~\cite{ATL-PHYS-PUB-2017-005} and single boson
($W/Z$+jets)~\cite{ATL-PHYS-PUB-2016-003} production, as well as electroweak
production of $W^\pm W^\pm+2$ jets, $WZ+2$ jets, and $ZZ+2$ jets, were modelled
using \SHERPAV{2.2.2} with the \pdf{NNPDF3.0NNLO} PDF set.  In order to improve
the agreement between the simulated and observed jet multiplicity distributions
for the $WZ \rightarrow \ell \nu \ell \ell$ and $ZZ \rightarrow \ell \ell \ell
\ell$ events, a jet-multiplicity based reweighting was applied to the simulated
$WZ$ and $ZZ$ samples.  Top-quark pair events (\ttbar) were generated using
\POWHEGBOXV{2}~\cite{Frixione:2007nw} interfaced to
\PYTHIAV{8.230}~\cite{pythia} and \textsc{EvtGen}~1.6.0. The
\pdf{NNPDF3.0NLO} PDF set was used for the matrix-element calculation, while
the \pdf{NNPDF2.3LO} PDF set was used for the showering with the \tune{A14} set
of tuned parameters.  Other background processes containing top quarks were
generated with \MGMCatNLO~\cite{amcnlo} interfaced to \PYTHIAV{8}, at LO
($\ttbar\gamma$, $tZ$, $t\bar{t}WW$, and $t\bar{t}t\bar{t}$) or at NLO (\ttW,
\ttZ, and \ttH), with \MGMCatNLO interfaced to \Herwig~\cite{Bahr:2008pv}
($tWZ$ and $tWH$) or with \POWHEGBOXV{2}~\cite{Re:2010bp} interfaced to
\PYTHIAV{6} ($tW$).

\section{Object definitions and selection criteria}
\label{s:selection}

Selected events are required to contain at least one reconstructed primary
vertex.  If more than one vertex is found, the vertex with the largest $\pT^2$
sum of associated ID tracks is selected as the primary vertex. 

Electrons are reconstructed as energy clusters in the EM calorimeter that are
matched to tracks found in the ID.  Muons are reconstructed by combining tracks
reconstructed in the ID with tracks or track segments found in the MS.  Leptons
need to satisfy $\pt > \SI{15}{\GeV}$ and have $|\eta|<2.47$ for electrons
(electrons within the transition region between the barrel and endcap
calorimeters, $1.37<|\eta|<1.52$, are excluded) and $|\eta|<2.5$ for muons.
Leptons are required to be consistent with originating from the primary vertex
by imposing requirements on the transverse impact parameter, $d_0$, its
uncertainty, $\sigma_{d_0}$, the longitudinal impact parameter, $z_0$, and the
polar angle $\theta$.  These requirements are $|d_{0}|/\sigma_{d_0}<5$ and
$|z_{0} \times \sin\theta| < \SI{0.5}{mm}$ for electrons, and
$|d_{0}|/\sigma_{d_0}<3$ and $|z_{0} \times \sin\theta| < \SI{0.5}{mm}$ for
muons.  
Electrons have to satisfy the likelihood-based ``Tight'' quality definition
which results in efficiencies of $58\%$ at $\ET=\SI{4.5}{GeV}$ to $88\%$ at
$\ET=\SI{100}{\GeV}$~\cite{EGAM-2018-01}.  For the $WWW$ ($WVZ$) analysis,
muons are required to pass the ``Medium'' (``Loose'') identification criteria
which results in efficiencies of approximately $96\%$ ($98\%$) for muons from a
$Z \to \mu\mu$ sample~\cite{PERF-2015-10}. 

To reject jets misidentified as leptons or leptons from hadron decays
(including $b$- and $c$-hadron decays), referred to as ``non-prompt'' leptons
in the following, leptons are required to be isolated from other particles in
both the calorimeters and the ID.  The lepton isolation cone size is at most
$\Delta R = 0.2$, except for the muon isolation in the ID, where it is at most
$\Delta R = 0.3$.
Electrons are required to pass the ``Fix (Loose)'' isolation
requirement~\cite{PERF-2017-01} and muons are required to pass the ``Gradient''
(``FixedCutLoose'') isolation requirement~\cite{PERF-2015-10} for the $WWW$
($WVZ$) analysis.  The identification and isolation requirements for muons are
more restrictive in the \WWW analysis because a larger contamination from
non-prompt leptons is expected.  The electron Fix (Loose) isolation requirement
results in an efficiency above 95\%~\cite{EGAM-2018-01}.  The muon isolation
efficiency is above 90\% (99\%) for the Gradient isolation criteria for muons
with \pT of \SI{25}{\GeV} (\SI{60}{\GeV}), and the FixedCutLoose efficiency is
above 95\%~\cite{PERF-2015-10}.  

A dedicated boosted decision tree (BDT), termed ``\plt''~\cite{HIGG-2017-02},
is used to reject leptons likely to originate from heavy-flavour decays.  In
addition, electrons have to pass the ``\cft''~\cite{PERF-2017-01} to reject
electrons likely to have the electric charge wrongly measured. 
The \plt\ uses isolation and $b$-tagging information derived from energy
deposits and tracks in a cone around the lepton direction.  The \cft\ uses the
electron track impact parameter, the track curvature significance, the cluster
width and the quality of the matching between the cluster and its associated
track.  Leptons passing all requirements listed above are referred to as
``nominal'' leptons. 
The combination of isolation, non-prompt lepton BDT and charge
misidentification suppression BDT criteria results in an efficiency of 65\%
(89\%) for electrons with \pT of \SI{20}{\GeV}
(\SI{80}{\GeV})~\cite{EGAM-2018-01}.  The combination of isolation and
non-prompt lepton BDT results in an efficiency of 65\% (99\%) for muons that
have the Gradient isolation with \pT of \SI{20}{\GeV} (\SI{100}{\GeV}), and an
efficiency of 74\% (99\%) for muons that have the FixedCutLoose isolation with
\pT of \SI{20}{\GeV} (\SI{80}{\GeV})~\cite{PERF-2015-10}.

Jets are reconstructed from calibrated topological clusters built from energy
deposits in the calorimeter~\cite{PERF-2014-07} using the anti-$k_t$ algorithm
with a radius parameter of 0.4~\cite{Cacciari:2008gp, Cacciari:2005hq} and
calibrated using the techniques described in Ref.~\cite{PERF-2016-04}.  Jet
candidates are required to have $\pt > \SI{20}{\GeV}$ and $|\eta| < 2.5$. To
reject jets likely to be arising from pile-up collisions, an additional
criterion using the jet vertex tagger~\cite{ATLAS-CONF-2014-018} discriminant
is applied for jets with $\pt < \SI{60}{\GeV}$ and $|\eta| < 2.4$.  Jets
containing $b$-hadrons ($b$-jets) are identified by a multivariate discriminant
combining information from algorithms using secondary vertices reconstructed
within the jet and track impact parameters~\cite{ATL-PHYS-PUB-2015-022,
ATL-PHYS-PUB-2016-012}, with an efficiency of 85\% (70\%) for the $WWW$ (\WVZ)
analysis.

The missing transverse momentum, whose magnitude is denoted \met, is defined as
the negative vector sum of the \pt of all reconstructed and calibrated objects
in the event. This sum includes a term to account for the energy from
low-momentum particles that are not associated with any of the selected
objects, and is calculated from ID tracks matched to the reconstructed primary
vertex in the event~\cite{PERF-2016-07}.  The sum also includes jets with
$|\eta|>2.5$ and $\pt > \SI{30}{\GeV}$.

The object reconstruction and identification algorithms do not always result in
unambiguous identifications. An overlap removal algorithm is therefore applied.
Electrons sharing a track with any muons are removed.  Any jet within $\Delta R
< 0.2$ of an electron is removed and electrons within $\Delta R < 0.4$ of any
remaining jets are removed.  Jets with less than three associated tracks and
within $\Delta R < 0.2$ of a muon are removed, and muons within $\Delta R  <
0.4$ of any of the remaining jets are removed.

At least one reconstructed ``trigger'' lepton with a minimum \pT is required to
match within $\Delta R < 0.15$ a lepton with the same flavour reconstructed by
the trigger algorithm.  The thresholds for the trigger (other) leptons are
\SI{27}{\GeV} (\SI{20}{\GeV}) for the \WWW analysis, and from \SI{21}{\GeV} to
\SI{27}{\GeV} (\SI{15}{\GeV}), depending on the run period and lepton flavour,
for the \WVZ analysis. 

\section{Analysis targeting \WWW}
\label{s:www}

The experimental signature of the \lljj process is the presence of two
same-sign leptons, \met, and two jets. The signature of the \lllv process is
the presence of three leptons and \met.  To reduce the background contributions
from processes that have more than two (three) leptons in the \ssll (\threelep)
channel a ``veto lepton'' definition is introduced.  Compared with the nominal
lepton selection criteria described in Section~\ref{s:selection}, the veto
lepton \pt threshold is lowered to \SI{7}{\GeV}, and the isolation, \plt, \cft,
and impact parameter requirements are removed. For veto electrons, the
likelihood-based Loose identification definition~\cite{PERF-2017-01} is used.
For veto muons, the Loose identification definition~\cite{PERF-2015-10} is
used, and the pseudorapidity range is extended to $|\eta|<2.7$.

To select \ssll candidates, events are required to have exactly two nominal
leptons with $\pt > \SI{20}{\GeV}$ and the same electric charge, at least two
jets, and no identified $b$-jets. Four regions are considered, based on the
lepton flavour, namely \elel, \elmu, \muel, and \mum, where \elmu denotes the
highest-\pt (leading) lepton being an electron, while \muel denotes the leading
lepton being a muon.  Events with an additional veto lepton are removed.  The
invariant mass of the dilepton system is required to be in the range $40 <
m_{\ell\ell} < \SI{400}{\GeV}$.  The upper mass limit reduces the contribution
from the $WZ$+jets process.  The leading (sub-leading) jet must have $\pt >
30\,(20)\,\GeV$ and $|\eta| < 2.5$. 
The dijet system, formed by the two jets with the largest \pT, is required to have
$m_{jj} < \SI{300}{\GeV}$ and $|\Delta \eta_{jj}| < 1.5$, where $m_{jj}$ is the
dijet invariant mass and $\Delta \eta_{jj}$ is the pseudorapidity separation
between the two jets.
The cuts applied on the dijet system mainly reduce the
contributions from the same-sign $WW$ vector boson scattering process.
Additionally, in the \elel final state, \met is required to be above
\SI{55}{\GeV} and $m_{\ell\ell}$ must satisfy $m_{\ell\ell} < \SI{80}{\GeV}$ or $m_{\ell\ell} > \SI{100}{\GeV}$,
to reduce contamination from $Z \to ee$ where the charge of one
electron is misidentified.  This $m_{\ell \ell}$ cut is not applied in the
$\mu\mu$ final state, since the muon charge misidentification rate is found to
be negligible, nor is it applied in the $e\mu$ and $\mu e$ final states, where
the contamination from $Z$ events is small.

To select \threelep candidates, events are required to have exactly three
nominal leptons with $\pt > \SI{20}{\GeV}$ and no identified $b$-jets.  Events
with an additional veto lepton are removed.  To reduce the contribution from
the $WZ \to \ell \nu \ell \ell$ process, events are required to have no
same-flavour opposite-sign (SFOS) lepton pairs, and thus only $\mu^\pm e^\mp
e^\mp$ and $e^\pm \mu^\mp \mu^\mp$ events are selected.

\begin{figure}[htbp]
\centering
\subfloat[]{\includegraphics[width=.48\textwidth]{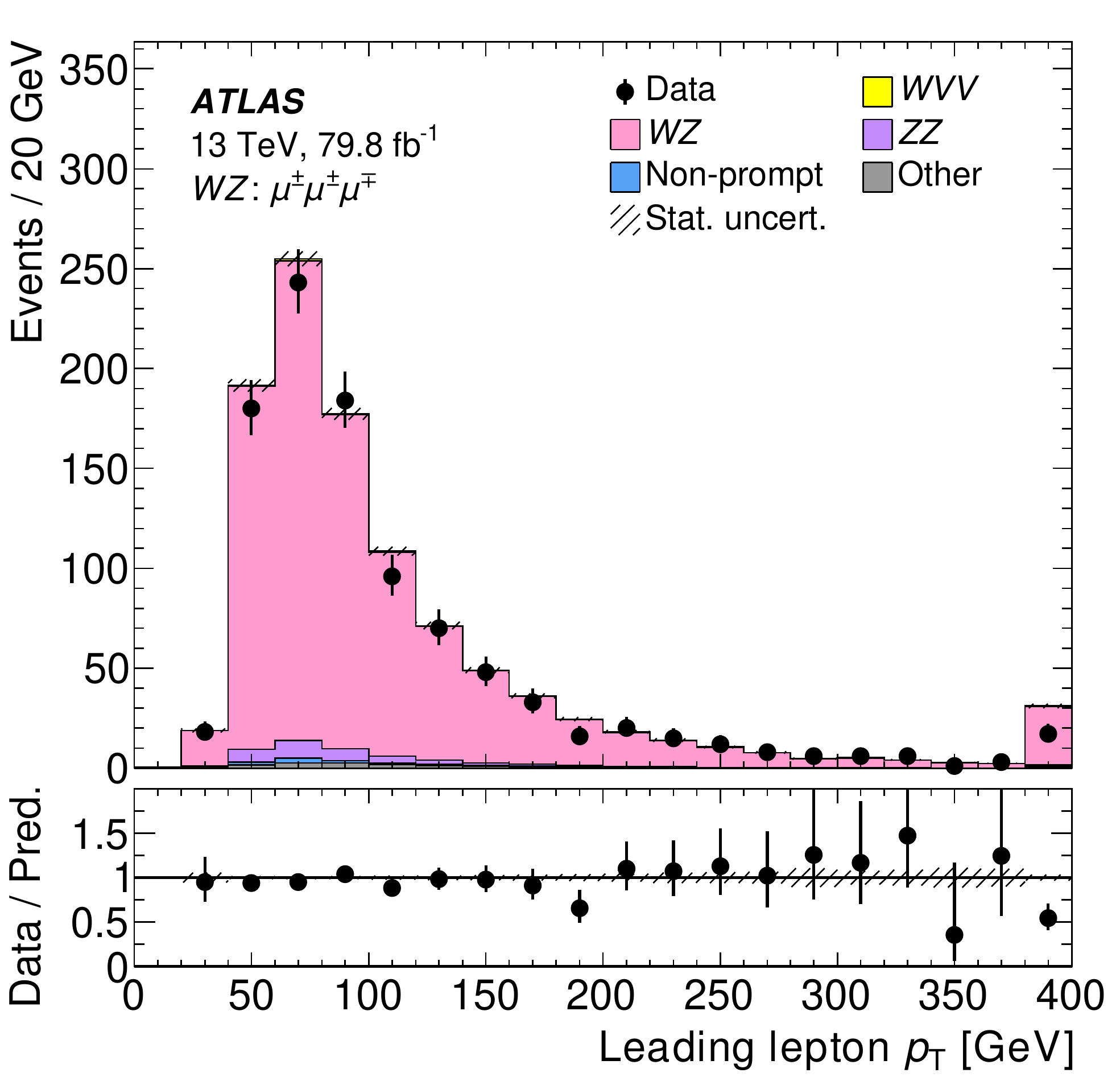}}
\hfill
\subfloat[]{\includegraphics[width=.48\textwidth]{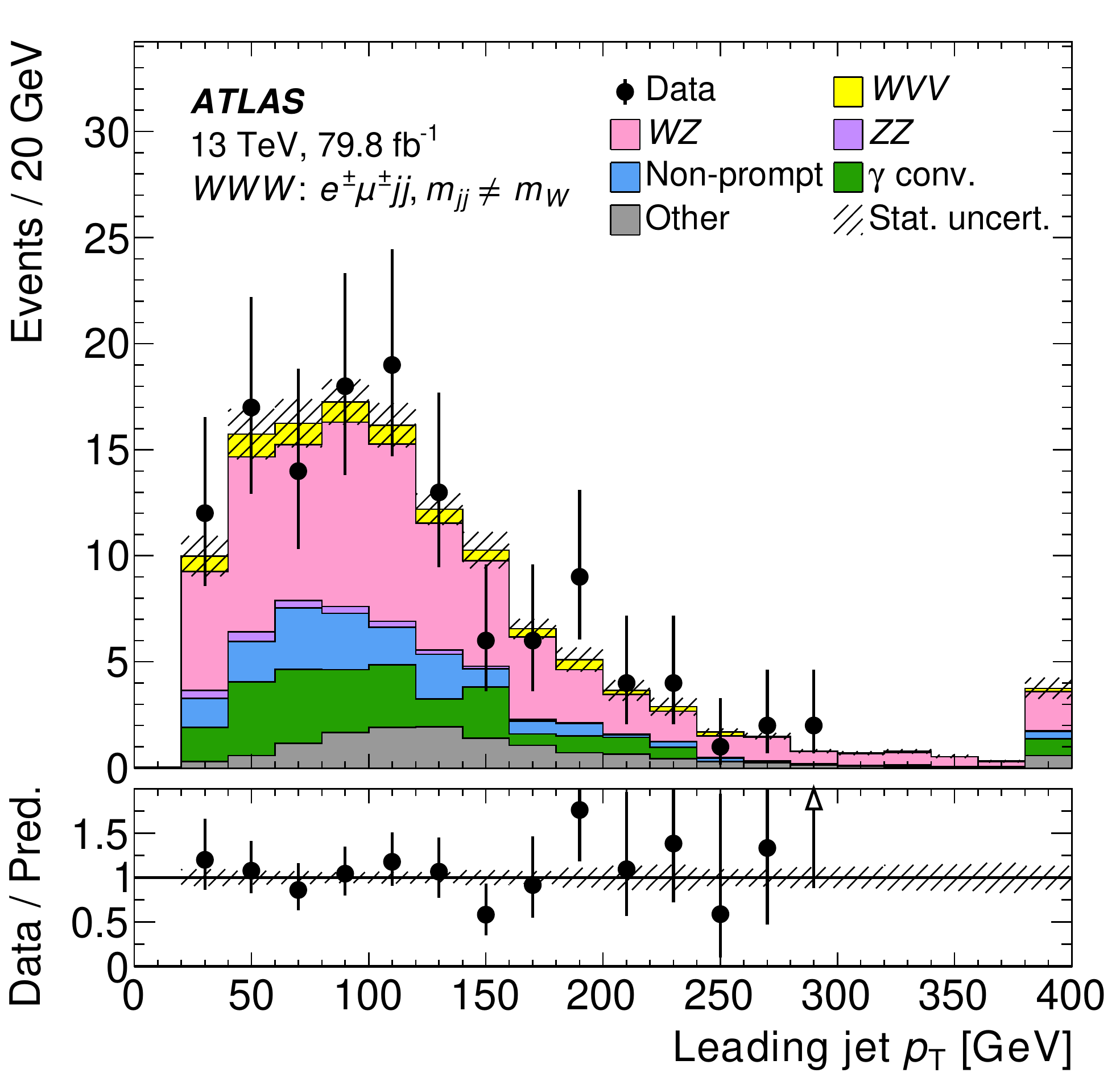}}
\caption{Comparison between data and prediction of (a) the leading lepton \pt
distribution in the $WZ$ validation region and (b) the leading jet \pt
distribution in the $W$ sideband region.  The contribution denoted ``Other'' is
dominated by the $W^\pm W^\pm+2$ jets process for the $W$ sideband region.  The
contribution denoted ``$\gamma$-conv.'' is described in the text.  \predscale.
\statband.  \oflo. \ratiopan.} 
\label{fig:www_validation}
\end{figure}

A major background originates from the $WZ$+jets $\to \ell \nu \ell \ell$+jets
process, contributing to the \ssll channel when one lepton is not reconstructed
or identified, or to the \threelep channel, when a $Z$ boson decays into a pair
of $\tau$ leptons both of which decay to an electron or muon.  Simulation is
used to estimate this background.  The $WZ$+jets modelling is tested in a
$WZ$-dominated validation region defined by selecting events with exactly three
nominal leptons with one SFOS lepton pair.  In addition, events are required to
have no $b$-jets reconstructed, $\met > \SI{55}{\GeV}$ and the trilepton
invariant mass $m_{\ell\ell\ell} > \SI{110}{\GeV}$.  Data and simulation agree
in this validation region, as shown in Figure~\ref{fig:www_validation}(a) for
the leading lepton \pt distribution.

Contributions from SM processes that produce at least one non-prompt lepton are
estimated using a data-driven method as described in Ref.~\cite{STDM-2014-05}
by introducing ``fake'' leptons.  The definitions of nominal and fake leptons
are mutually exclusive. Fake electrons have to satisfy the likelihood-based
Medium~\cite{PERF-2017-01} but fail the Tight identification, and the
isolation, \pltcft\ are removed.  Fake muons have the impact parameter
requirements loosened to $|d_{0}|/\sigma_{d_0}<10$, and both isolation and
\plt\ requirements are removed. Additionally, they have to fail the nominal
muon definition.  Simulation shows that the \ttbar process is the dominant
contributor of events with fake leptons, with more than 90\% in the \ssll
channel and more than 95\% in the \threelep channel originating from this
process.  Events containing one (two) nominal lepton(s) and one fake lepton
with $\pt > \SI{20}{\GeV}$ are scaled by a ``fake factor'' to predict the
non-prompt lepton background contribution in the \ssll (\threelep) channel.
The fake factor is the ratio of the number of non-prompt leptons passing the
nominal lepton criteria over the number passing the fake lepton criteria. Its
value is derived from two \ttbar-enriched regions selected with two or three
leptons (no SFOS lepton pairs) and exactly one $b$-jet. 
One of the same-sign
leptons passes either nominal or fake lepton criteria, while the other
lepton(s) must pass the nominal lepton criteria.
The fake factor is found to be 
$0.017 \pm 0.010$ for electrons and $0.035 \pm 0.005$ for muons.  

Events resulting from the $V\gamma jj$ production can pass the \elel, \elmu or
\muel signal selection criteria if the photon is misreconstructed as an
electron.  This contribution (referred to as ``$\gamma$ conv.'') is evaluated
using a data-driven method similar to the non-prompt lepton background
evaluation by introducing ``photon-like'' electrons.  A photon-like electron is
an object reconstructed like a nominal electron except that the track has no
hit in the innermost layer of the pixel detector and the \pltcft\ are not
applied.  The photon fake factor is determined in two regions selected with two
nominal muons, no $b$-jets, and one nominal or photon-like electron.  The
trilepton invariant mass is required to satisfy 
$\SI{80}{\GeV} < m_{e \mu\mu} < \SI{100}{\GeV}$.
Most of these events contain a $Z \rightarrow \mu \mu$
decay, where one muon radiates a photon, which is misreconstructed as an
electron. 

The charge misidentification background originates from processes that produce
oppositely-charged prompt leptons, where one lepton's charge is misidentified
and results in final states with two same-sign leptons.  The background is
estimated using a data-driven technique as described in
Ref.~\cite{STDM-2015-07}.  

All \ssll candidates with $m_{jj} < \SI{50}{\GeV}$ or $m_{jj} > \SI{120}{\GeV}$ (denoted
as the ``$W$ sideband'' region) are used to validate the modelling of different
backgrounds described above. Data and prediction agree, as shown in
Figure~\ref{fig:www_validation}(b) for the leading jet \pt distribution. 
Events with $m_{jj} < \SI{300}{\GeV}$ are used in the fit to extract the signal.

\section{Analysis targeting \WWZ and \WZZ}
\label{s:wvz}

The experimental signature of the \lllwvz, \llllwwz, and \llllwzz processes is
the presence of three or four charged leptons.  In order to increase the signal
acceptance, ``loose'' leptons are defined in addition to nominal leptons, the
latter being a subset of the former.  Loose leptons have both the isolation and
\plt\ requirements removed.  In addition, loose electrons are required to pass
the likelihood-based Loose identification definition and the \cft\ requirement
is removed.

Six regions are defined with either three or four loose leptons, sensitive to
triboson final states containing $Z$ bosons.  Among all possible SFOS lepton
pairs, the one with $m_{\ell\ell}$ closest to the $Z$ boson mass is defined as the best $Z$
candidate.  In all regions, the presence of such a best $Z$ candidate with
$|m_{\ell\ell} - \SI{91.2}{\GeV}| < \SI{10}{\GeV}$, is required.  Furthermore, any SFOS
lepton pair combination is required to have a minimum invariant mass of
$m_{\ell\ell} > \SI{12}{\GeV}$.  Events with $b$-tagged jets are vetoed.

For the three-lepton channel, the lepton which is not part of the best $Z$ candidate
is required to be a nominal lepton. The scalar sum of the transverse momenta of
all leptons and jets (\HT) is required to be larger than \SI{200}{\GeV}. This
significantly reduces the contribution of the $Z\to\ell\ell$ processes with one
additional non-prompt lepton.  Three regions are defined according to the
number of jets in the event: one jet (\TLjone), two jets (\TLjtwo), and at
least three jets (\TLjthree).

For the four-lepton channel, the third and fourth leading leptons are required
to be nominal leptons. The two leptons which are not part of the best $Z$ candidate
definition are required to have opposite charges. These ``other leptons'' are used to define three
regions, depending on whether they are different-flavour (\FLDF), or
same-flavour and their mass lies within a window of \SI{10}{\GeV} around the
$Z$ boson mass (\FLSFZ) or their mass is outside this window (\FLSFnoZ).

In each of the six regions the distribution of a dedicated
BDT discriminant, separating the \WVZ signal from the
dominating diboson background, is fed as input to the binned maximum-likelihood
fit to extract the signal.  
For the three-lepton channels, 13, 15, and 12 input
variables are used for the \TLjone, \TLjtwo, and \TLjthree final states,
respectively, while for the four-lepton channels, six input
variables are used for each of the \FLDF, \FLSFZ and \FLSFnoZ final states.
These input variables are listed in Table~\ref{tab:variables}.

\begin{table}[htbp]
\caption{List of input variables used in the multivariate analysis for each of the 
$WVZ$ channels, denoted by $\times$. The subscripts $1$, $2$ and $3$ refer to the
leading, subleading and third leading lepton or jet.  The definitions of ``best
$Z$ candidate'' and ``other leptons'' are given in the text. The variable
$m_{\text{T}}(W_{\ell})$ is the $W$-boson transverse mass of the leptonically decaying
$W$-boson candidate. 
Among the invariant masses formed by all possible jet pairs, the one closest to the
$W$-boson mass defines the ``$m_{jj}$ of best $W$ candidate'' and the smallest one defines
the ``smallest $m_{jj}$''.
Finally, the leptonic and hadronic \HT are calculated as the scalar sum of the \pT of
all leptons or all jets, respectively.
}
\label{tab:variables}
\begin{center}
\begin{tabular}{p{3.8cm}*{6}{c}}
\toprule \toprule
Variable & \TLjone & \TLjtwo & \TLjthree & $4\ell$ DF & $4\ell$ SF & $4\ell$ SF\\
& \phantom{off-shell} & \phantom{off-shell} & \phantom{off-shell} & \phantom{off-shell} & on-shell & off-shell\\
\midrule
$\pT(\ell_1)$		& $\times$ & $\times$ &  &  &  & \\
$\pT(\ell_2)$		& $\times$ & $\times$ & $\times$ &  &  & \\
$\pT(\ell_3)$		& $\times$ & $\times$ & $\times$ &  &  & \\
Sum of $\pT(\ell)$      & $\times$ & $\times$ & $\times$ &  &  & \\
$m_{\ell_1 \ell_2}$	& $\times$ & $\times$ &  &  &  & \\
$m_{\ell_1 \ell_3}$	& $\times$ & $\times$ &  &  &  & \\
$m_{\ell_2 \ell_3}$	& $\times$ & $\times$ &  &  &  & \\
$m_{\ell\ell}$ of best $Z$ candidate &  &  &  &  & $\times$ & $\times$\\
$m_{\ell\ell}$ of other leptons &  &  &  & $\times$ & $\times$ & $\times$\\
$m_{3\ell}$		& $\times$ & $\times$ & $\times$ & & & \\
$m_{4\ell}$ &  &  &  & $\times$ & $\times$ & $\times$\\
Sum of lepton charges	& $\times$ & $\times$ & $\times$ &  &  & \\
$\pT(j_1)$ & $\times$ & $\times$ &  &  &  & \\ 
$\pT(j_2)$ &  & $\times$ & $\times$ &  &  & \\
Sum of $\pT(j)$         &  &  & $\times$ &  &  & \\
Number of jets&  &  & $\times$ & $\times$ & $\times$ & $\times$\\
$m_{j_1 j_2}$ &  & $\times$ &  &  &  & \\
$m_{\text{T}}(W_{\ell}) $ &  & $\times$ &  &  &  & \\
$m_{jj}$ of best $W$ candidate &  &  & $\times$ &  &  & \\
Smallest $m_{jj}$	&  &  & $\times$ &  &  & \\
\met			&  & $\times$ & $\times$ & $\times$ & $\times$ & $\times$\\
\HT 		& $\times$ & $\times$ &  &  & $\times$ & $\times$\\
Leptonic \HT                  &  &  &  & $\times$ &  & \\
Hadronic \HT                  &  &  &  & $\times$ &  & \\
Invariant mass of all\\
~~~~leptons, jets and \met & $\times$ &  & $\times$ &  &  & \\
Invariant mass of the\\ 
~~~~best $Z$ leptons and $j_1$ & $\times$ &  &  &  &  & \\
\bottomrule \bottomrule
\end{tabular}
\end{center}
\end{table}

\begin{figure}[htbp]
\centering
\subfloat[]{\includegraphics[width=.48\textwidth]{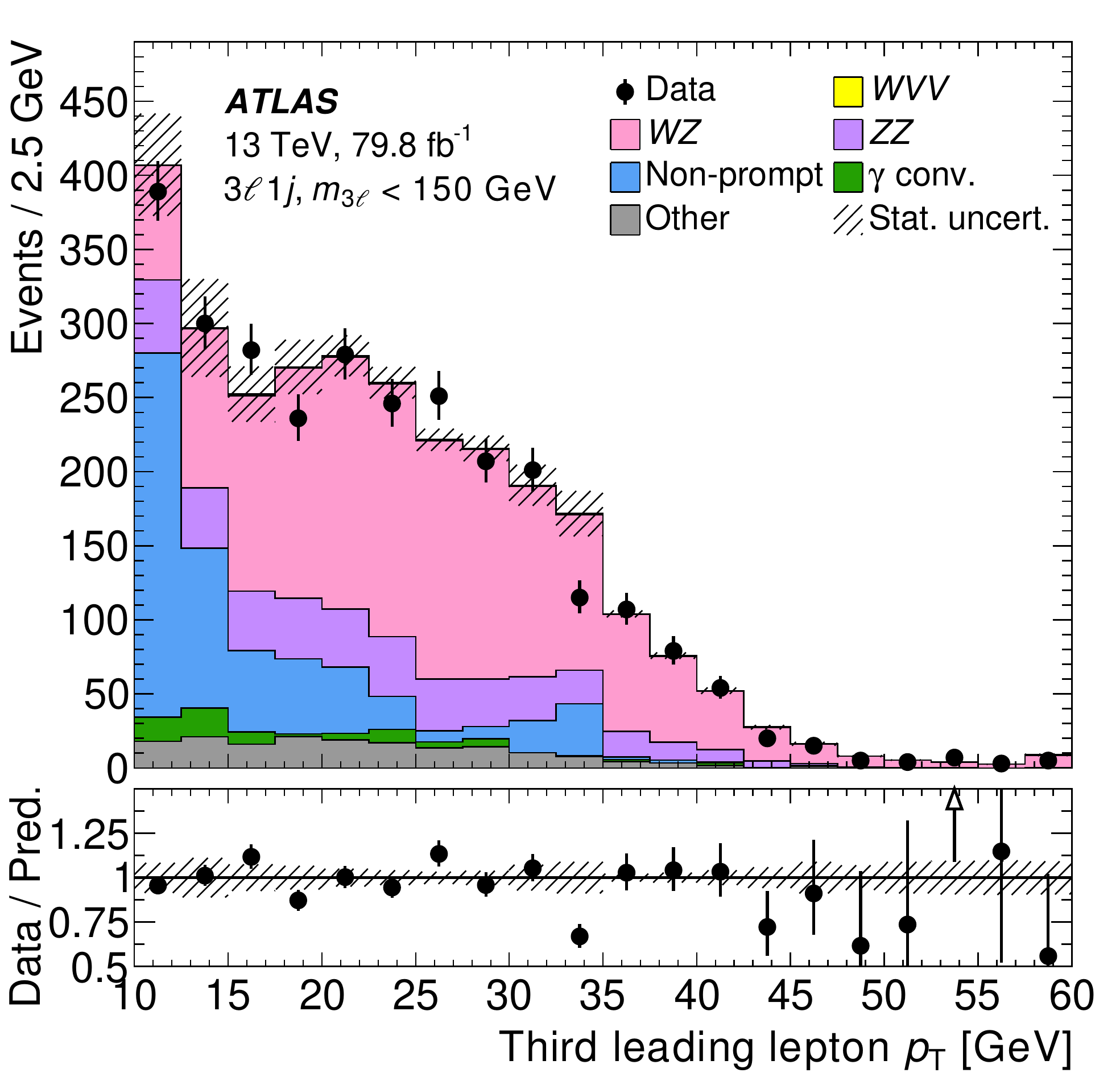}}
\subfloat[]{\includegraphics[width=.48\textwidth]{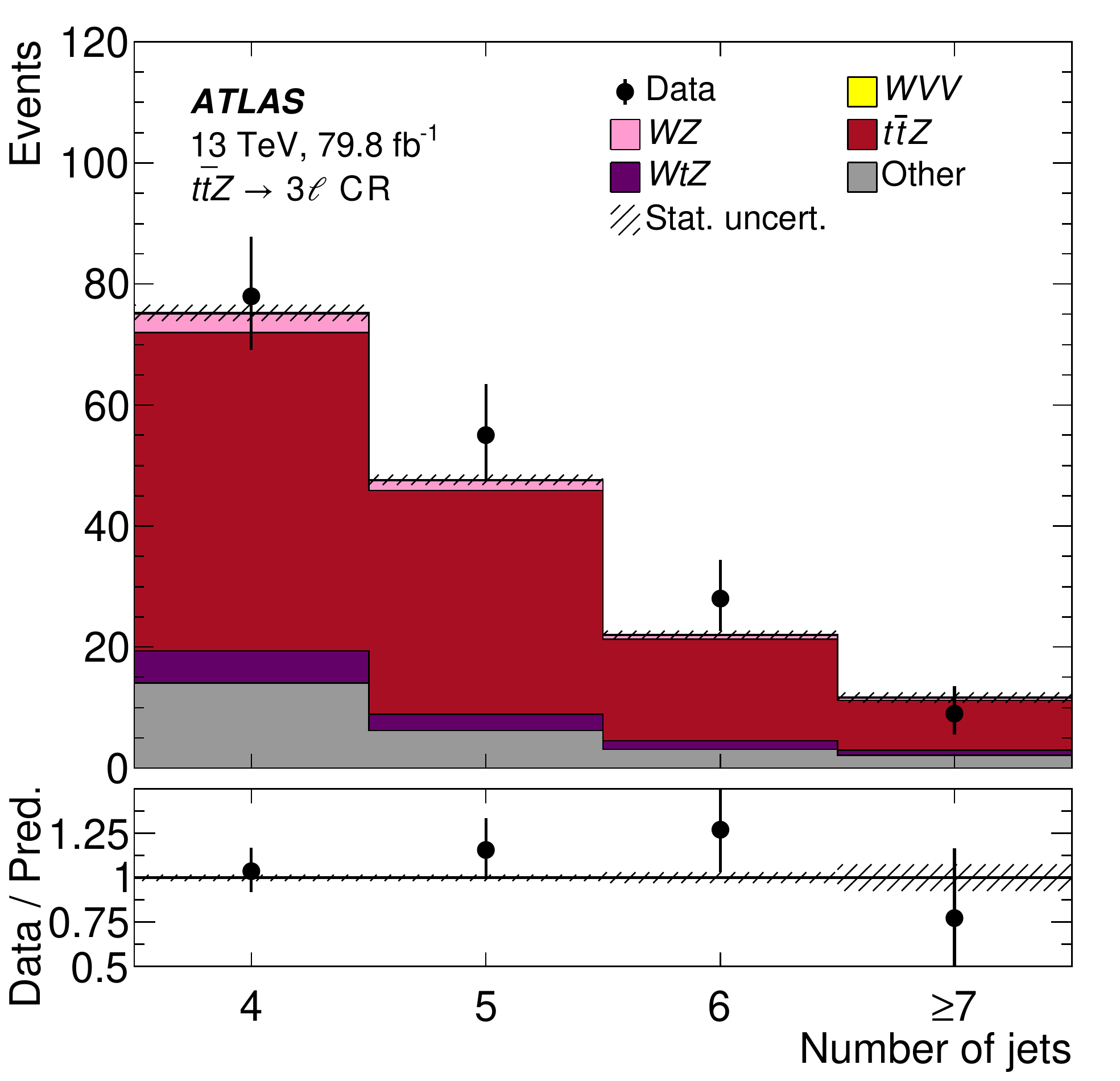}}
\caption{Data compared with expectations for (a) the transverse momentum of the
third-highest-\pt lepton in the \TLjone region with the additional requirement
$m_{\ell\ell\ell} < \SI{150}{\GeV}$, no requirement on \HT, and including the
$\SI{10}{\GeV} < \pt < \SI{15}{\GeV}$ validation region and (b) the number of
jets in the \ttZ control region.  The contributions denoted ``Other'' are
dominated by (a) the $tZ$ and $VH$ processes, where the Higgs boson does not
decay to two massive bosons, and (b) the $tZ$ process.  \predscale.  \statband.
\oflo.  \ratiopan.}
\label{fig:tlvalid}
\end{figure}

Due to the required presence of nominal leptons in the three- and four-lepton
channels, backgrounds with a $Z$ boson and non-prompt leptons are reduced.  The
remaining backgrounds are dominated by processes with prompt leptons and thus
all backgrounds are estimated using simulation.  The $WZ+$jets and $Z$+jets
backgrounds are validated in a region defined in the same way as the \TLjone
region, with the exception that no requirement on \HT is applied, the
third-highest-\pt lepton is required to have a small transverse momentum
($\SI{10}{\GeV} < \pt < \SI{15}{\GeV}$), and the invariant mass of the three
leptons has to be smaller than \SI{150}{\GeV}.  Data and expectation agree in
the \TLjone validation region, as shown in Figure~\ref{fig:tlvalid}(a) for the
transverse momentum distribution of the third-highest-\pt lepton. 

The \ttZ background is determined in a region defined like the \TLjthree region
with the exception that no requirement on \HT is applied, and at least four
jets are required, of which at least two are $b$-tagged. This region is
included as a single-bin control region (CR) in the fit model, outlined in
Section~\ref{s:results}.  Data and expectation agree, as shown in
Figure~\ref{fig:tlvalid}(b) for the \ttZ control region.

\section{Signal extraction and combination} 
\label{s:results}

The \WWW, \WWZ and \WZZ regions are combined using the profile likelihood
method described in Ref.~\cite{ATL-PHYS-PUB-2011-011} based on a simultaneous
fit to distributions in the signal and background control regions.
A total of eleven signal regions are considered: four regions (\elel, \elmu,
\muel, and \mum) for the \ssll channel, one region ($\mu e e$ and $e \mu \mu$
combined) for the \threelep channel, three regions (\TLjone, \TLjtwo, and
\TLjthree) for the \WVZ three-lepton channel, and three regions (\FLDF, \FLSFZ,
and \FLSFnoZ) for the \WVZ four-lepton channel.  One control region is
considered: the \ttZ control region described in Section~\ref{s:wvz}.  The
distributions used in the fit are the $m_{jj}$ distributions for the \ssll
channel and the BDT distributions for the \WVZ three-lepton and four-lepton
channels. The number of selected events in the \threelep channel and the \ttZ
control region are each included as a single bin in the fit. In total, 186 bins
are used in the combined fit. 

A binned likelihood function ${\cal{L}} (\mu$, {\textbf{$\theta$}}) is
constructed as a product of Poisson probability terms over all bins considered.
This likelihood function depends on the signal-strength parameter $\mu$, a
multiplicative factor that scales the number of expected signal events, and
{\textbf{$\theta$}}, a set of nuisance parameters that encode the effect of
systematic uncertainties in the signal and background expectations. The
nuisance parameters are implemented in the likelihood function as Gaussian,
log-normal or Poisson constraints.  The same value for $\mu = \mu_{WVV}$ 
is assumed for the
on- and off-mass-shell \WWW, \WWZ and \WZZ processes.  Correlations of
systematic uncertainties arising from common sources are maintained across
processes and channels.  

Experimental uncertainties are related to the lepton trigger, reconstruction
and identification efficiencies~\cite{PERF-2017-01, PERF-2015-10}, lepton
isolation criteria~\cite{HIGG-2017-02}, lepton energy (momentum) scale and
resolution~\cite{PERF-2015-10, PERF-2017-03}, jet energy scale and
resolution~\cite{PERF-2016-04}, jet vertex tagging~\cite{ATLAS-CONF-2014-018,
PERF-2016-06}, $b$-tagging~\cite{ATL-PHYS-PUB-2016-012}, modelling of pile-up
and missing transverse momentum~\cite{PERF-2016-07}, and integrated
luminosity~\cite{DAPR-2013-01,LUCID2}.  Nuisance parameters related to these
uncertainties are treated as correlated between all channels.
The time-dependence of the efficiencies, scales and resolutions across the 
various run periods is taken into account.

For each of the background processes evaluated using simulation, a nuisance
parameter representing its normalisation uncertainty is included.  The
following prior uncertainties in the normalisations are assumed: 20\% for $WZ$
and $ZZ$; 40\% for $Z$+jets, 10\%~\cite{TOPQ-2016-11} for $WtZ$,
30\%~\cite{TOPQ-2016-14, TOPQ-2017-06} for $tZ$, 11\%~\cite{deFlorian:2016spz}
for \ttZ, and 30\% for $VH$ not producing three massive bosons. For dominant
backgrounds from the $WZ$ and $ZZ$ processes, the simultaneous fit model has
the power to constrain their normalisations at the $\sim$5\% level,
independently of the assumed prior.  In addition, shape-only variations for
backgrounds from the $WZ$ and $ZZ$ processes are derived from alternative
samples, generated using \POWHEG~\cite{Nason:2013ydw} with \PYTHIAV{8} for the
parton shower to account for differences in the modelling of diboson production
and showering.  Shape variations due to renormalisation and factorisation
scales are also considered for these two processes. The prior uncertainties
assumed for $Z$+jets and $VH$ cover the observed data/simulation agreement in
validation regions, and the calculations in Ref.~\cite{deFlorian:2016spz},
respectively. The impact of these uncertainties on the measurement is small.

Uncertainties in data-driven background evaluations mainly come from
statistical and systematic uncertainties in the charge misidentification rate,
lepton fake factor, and photon-like electron scale factor. Additional
uncertainties come from the statistical uncertainties in the subsamples used to
extrapolate the background evaluations to the signal region. Nuisance
parameters are treated as correlated for backgrounds evaluated using the same
method and from the same systematic sources. 

Shape-only variations of the signal distributions due to QCD renormalisation and
factorisation scales, PDF, and parton-shower matching scales are considered
in the simultaneous fit.  The corresponding nuisance parameters are treated as
correlated between the \lljj and \lllv channels in the \WWW analysis and
between three-lepton and four-lepton channels in the \WVZ analysis.  These
parameters are treated as uncorrelated between the \WWW and \WVZ analyses. 

Tables~\ref{tab:www_obs_postfit_yields} and \ref{tab:wvz_obs_postfit_yields}
show the post-fit background, signal and observed yields for the signal regions
and the background control region.
The contribution to the $WVV$ signal from $VH$ associated production is $\sim 40\%$ 
in the $WWW$ fiducial regions and $\sim 30\%$ in the $WVZ$ fiducial regions. 
Contributions from SM processes producing the same detector signature as events
in these signal regions (or the \ttZ control region) besides those listed are
combined into ``Other''.  The uncertainties shown include both statistical and
systematic uncertainties.  Data and predictions agree in all channels. 

\begin{table}[htbp]
\caption{Post-fit background, signal and observed yields for the 
\ssll and \threelep channels.
Uncertainties in the predictions include both statistical
and systematic uncertainties added in quadrature; correlations among systematic
uncertainties are taken into account in the calculation of the total.
\label{tab:www_obs_postfit_yields}}
\small
\begin{center}
\begin{tabular}{l|
      r@{$\:\pm\:$}l
      r@{$\:\pm\:$}l
      r@{$\:\pm\:$}l
      r@{$\:\pm\:$}l|
      r@{$\:\pm\:$}l
}
\toprule
\toprule
& \multicolumn{2}{c}{\elel} & \multicolumn{2}{c}{\elmu} &
\multicolumn{2}{c}{\muel} & \multicolumn{2}{c|}{\mum} & \multicolumn{2}{c}{$\mu
e e + e \mu \mu$}\\
\midrule
  $WWW$   & 9.9 & 3.3 & 26 & 9 & 23 & 8 & 30 & 10 & 15 & 5 \\ 
  $WZ$   & 37.4 & 2.2 & 121 & 6 & 96 & 5 & 119 & 6 & 8.6 & 0.5 \\ 
  $ZZ$   & 0.46 & 0.05 & 5.11 & 0.25 & 3.44 & 0.18 & 4.12 & 0.24 & 0.69 & 0.03 \\ 
  Non-prompt   & 6.1 & 3.0 & 35 & 5 & 17 & 9 & 37 & 7 & 9.4 & 1.5 \\ 
  $\gamma$ conv.   & 20.9 & 1.9 & 35.0 & 3.1 & 76 & 7 & \multicolumn{2}{c|}{--} & 1.06 & 0.11 \\ 
  Other   & 12.9 & 1.0 & 25.7 & 1.7 & 20.3 & 1.3 & 25.3 & 1.6 & 3.5 & 0.4\\ 
\midrule 
  Total  & 88 & 4 & 249 & 9 & 237 & 10 & 216 & 9 & 38 & 4 \\ 
  Data   & \multicolumn{2}{c}{87} & \multicolumn{2}{c}{239} &
\multicolumn{2}{c}{235} & \multicolumn{2}{c|}{237} & \multicolumn{2}{c}{27} \\ 
\bottomrule
\bottomrule
\end{tabular}
\end{center}
\end{table}

\begin{table}[htbp]
\caption{Post-fit background, signal and observed yields for the three-lepton
and four-lepton channels as well as the \ttZ control region.
Uncertainties in the predictions include both statistical and
systematic uncertainties added in quadrature; correlations among systematic
uncertainties are taken into account in the calculation of the total.
\label{tab:wvz_obs_postfit_yields}}
\small
\begin{center}
\begin{tabular}{l|
      r@{$\:\pm\:$}l
      r@{$\:\pm\:$}l
      r@{$\:\pm\:$}l|
      r@{$\:\pm\:$}l
      r@{$\:\pm\:$}l
      r@{$\:\pm\:$}l|
      r@{$\:\pm\:$}l
}
\toprule
\toprule
& \multicolumn{2}{c}{\FLDF}
& \multicolumn{2}{c}{\FLSFZ}
& \multicolumn{2}{c|}{\FLSFnoZ}
& \multicolumn{2}{c}{\TLjone}
& \multicolumn{2}{c}{\TLjtwo}
& \multicolumn{2}{c|}{\TLjthree}
& \multicolumn{2}{c}{\ttZ CR}\\
\midrule
  \WVZ   & 9.6 & 3.5 & 5.0 & 1.8 & 10 & 4 & 62 & 23 & 85 & 30 & 84 & 30 & \multicolumn{2}{c}{--} \\ 
  $WZ$   & 1.11 & 0.13 & \multicolumn{2}{c}{--} & 1.08 & 0.14 & 2580 & 80 & 1830 & 60 & 1110 & 50 & 5.7 & 0.4 \\ 
  $ZZ$   & 6.7 & 0.4 & 933 & 28 & 310 & 10 & 344 & 12 & 182 & 13 & 98 & 12 & 0.58 & 0.06 \\ 
  \ttZ   & 5.1 & 0.5 & 0.55 & 0.08 & 4.5 & 0.5 & 7.6 & 1.1 & 22.6 & 2.5 & 82 & 8 & 122 & 9 \\ 
  $tWZ$   & 1.9 & 0.4 & 0.23 & 0.10 & 1.6 & 0.4 & 4.2 & 0.9 & 11.2 & 2.2 & 20 & 4 & 10.3 & 0.8 \\ 
  Non-prompt   & \multicolumn{2}{c}{--} & \multicolumn{2}{c}{--} & 0.18 & 0.12 & 130 & 50 & 77 & 28 & 59 & 24 & 0.47 & 0.18 \\ 
  $\gamma$ conv.   & \multicolumn{2}{c}{--} & \multicolumn{2}{c}{--} & \multicolumn{2}{c|}{--} & 42 & 8 & 32 & 7 & 9.6 & 3.4 & 0.4 & 0.6 \\ 
  Other   & 0.4 & 0.4 & 1.8 & 1.1 & 1.0 & 0.7 & 200 & 15 & 182 & 16 & 120 & 10 & 24.4 & 2.5 \\ 
\midrule 
  Total  & 24.8 & 3.5 & 941 & 27 & 329 & 10 & 3370 & 70 & 2430 & 40 & 1580 & 40 & 160 & 10 \\ 
Data &   
\multicolumn{2}{c}{28} &
\multicolumn{2}{c}{912} &
\multicolumn{2}{c|}{360} & 
\multicolumn{2}{c}{\num{3351}} & 
\multicolumn{2}{c}{\num{2438}} & 
\multicolumn{2}{c|}{\num{1572}} & 
\multicolumn{2}{c}{170} \\
\bottomrule
\bottomrule
\end{tabular} 
\end{center}
\end{table}

\begin{figure}[htbp]
\centering
\subfloat[]{\includegraphics[width=.42\textwidth]{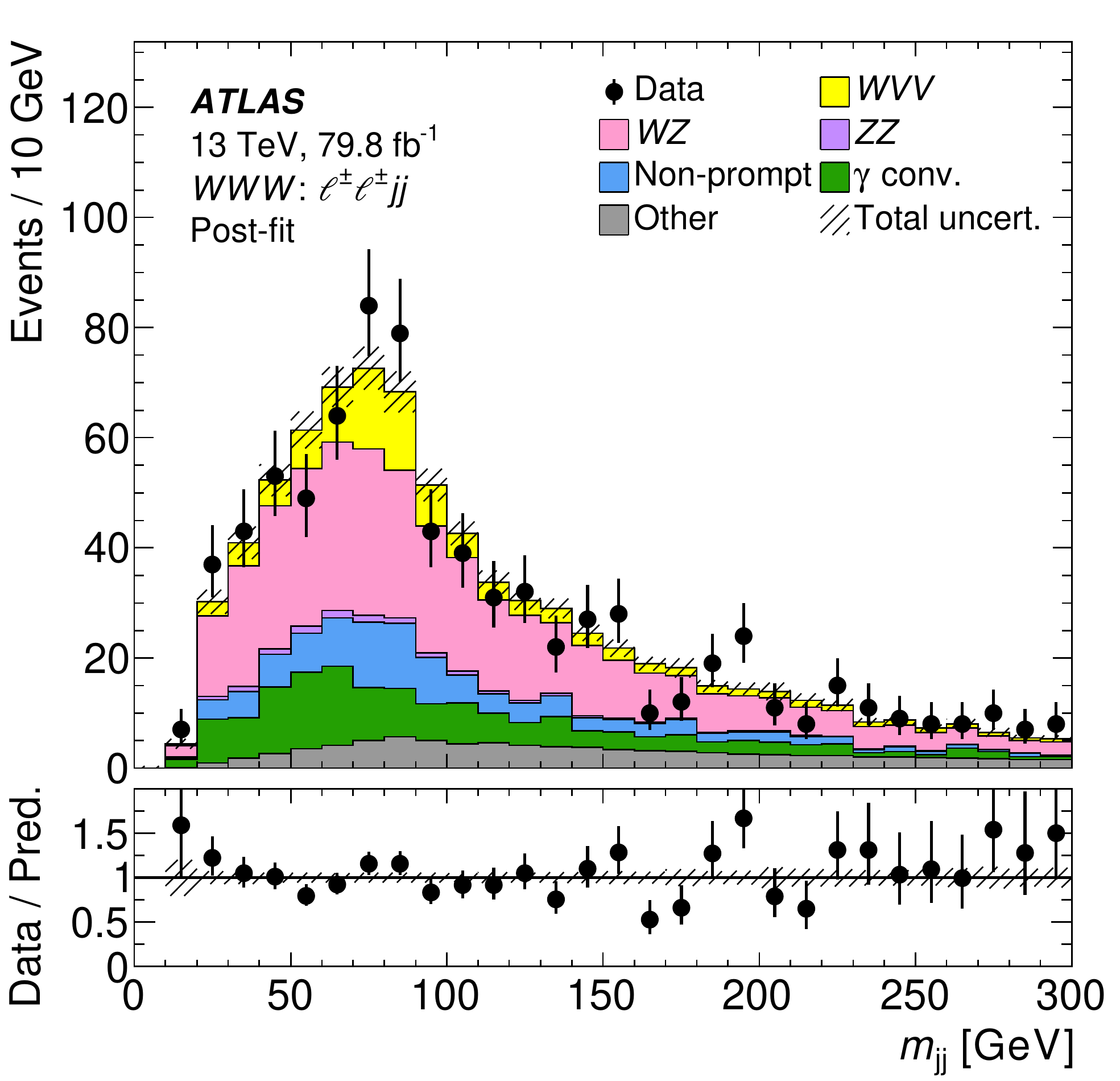}}
\subfloat[]{\includegraphics[width=.42\textwidth]{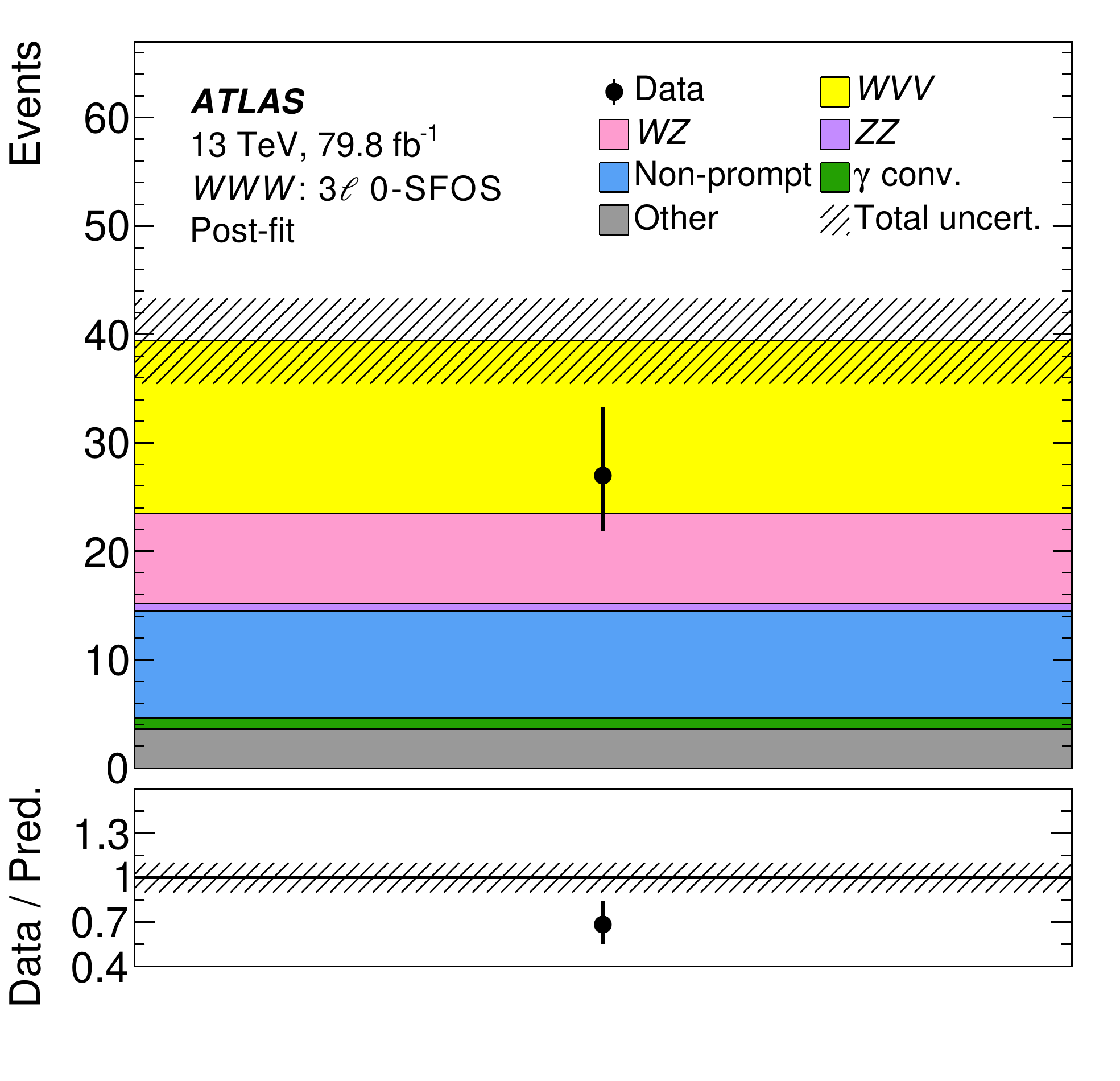}} \\
\subfloat[]{\includegraphics[width=.42\textwidth]{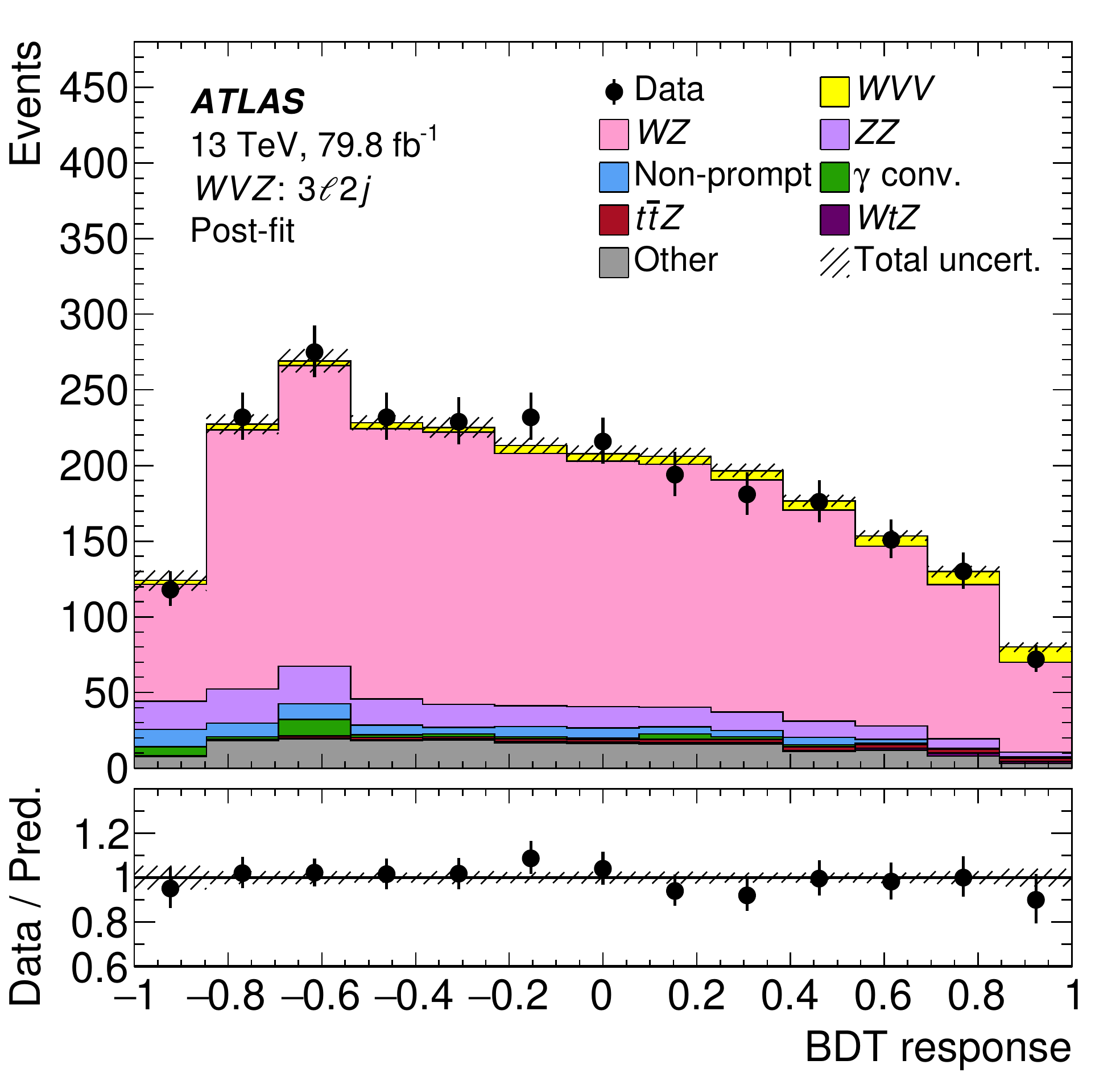}}
\subfloat[]{\includegraphics[width=.42\textwidth]{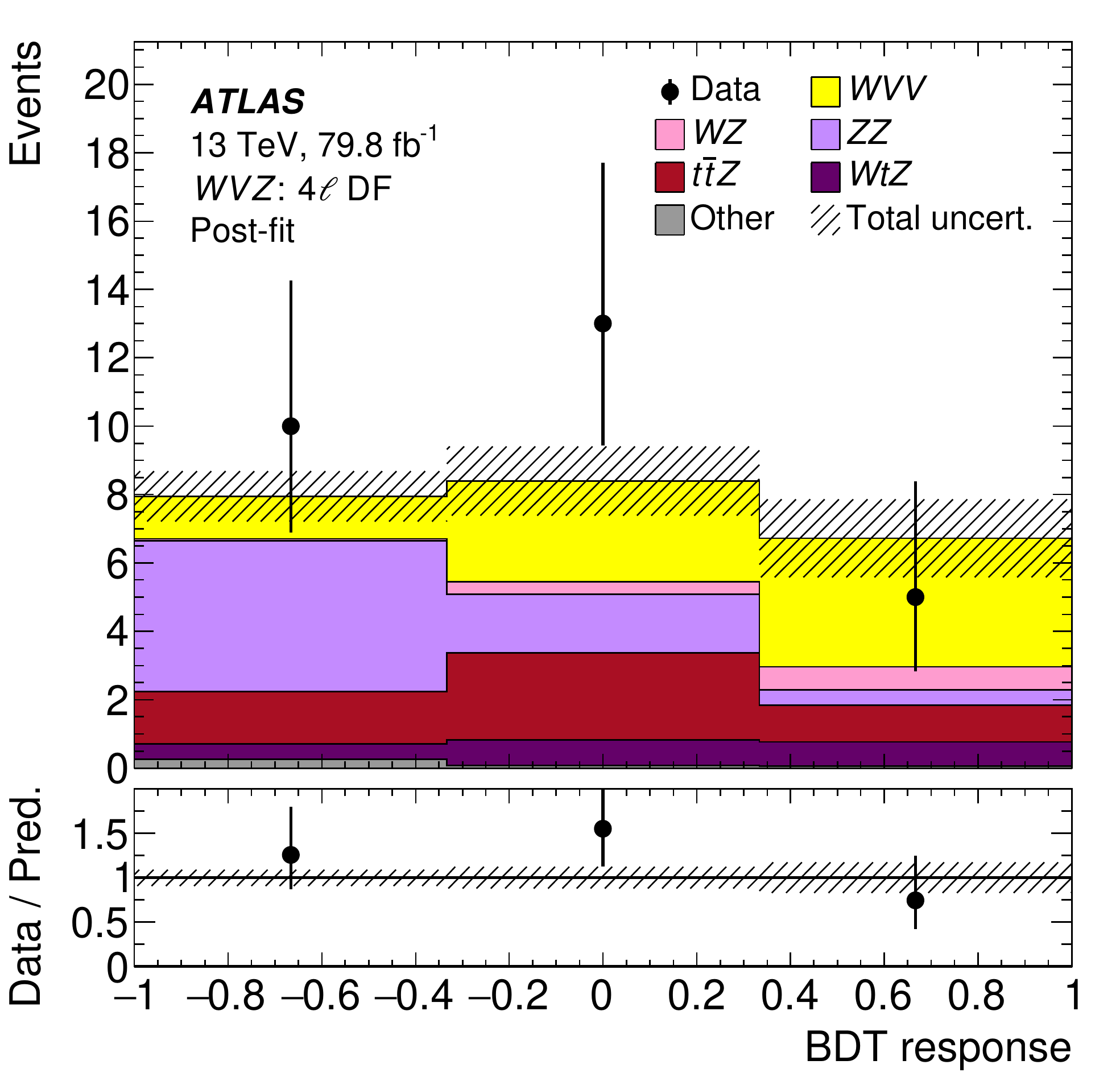}}
\caption{\label{fig:www_wvz_post} Post-fit distribution of (a) $m_{jj}$ for the
\sswww analysis (\elel, \elmu, \muel, \mum combined), (b) number of events for
the \lllwww analysis, and the BDT response in the (c) \TLjtwo and (d) \FLDF
channels for the $WVZ$ analysis.  The contributions denoted ``Other'' are
dominated by the (a) $W^{\pm}W^{\pm} +2$jets, (b) \ttW and (c) $tZ$ process,
respectively.  The uncertainty band includes both statistical and systematic
uncertainties as obtained by the fit.}
\end{figure}

Figure~\ref{fig:www_wvz_post} shows the comparison between data and post-fit
prediction of the combined $m_{jj}$ distribution for the \lljj channel, the
number of selected events for the \lllv channel, and the BDT output
distributions in the \TLjtwo and \FLDF regions for the \WVZ analysis.  The
\TLjtwo and \FLDF regions are chosen since they have the best sensitivity among
the three-lepton and four-lepton channels.  Data and predictions agree in all
distributions. 

\begin{figure}[htbp]
\centering
\includegraphics[width=.6\textwidth]{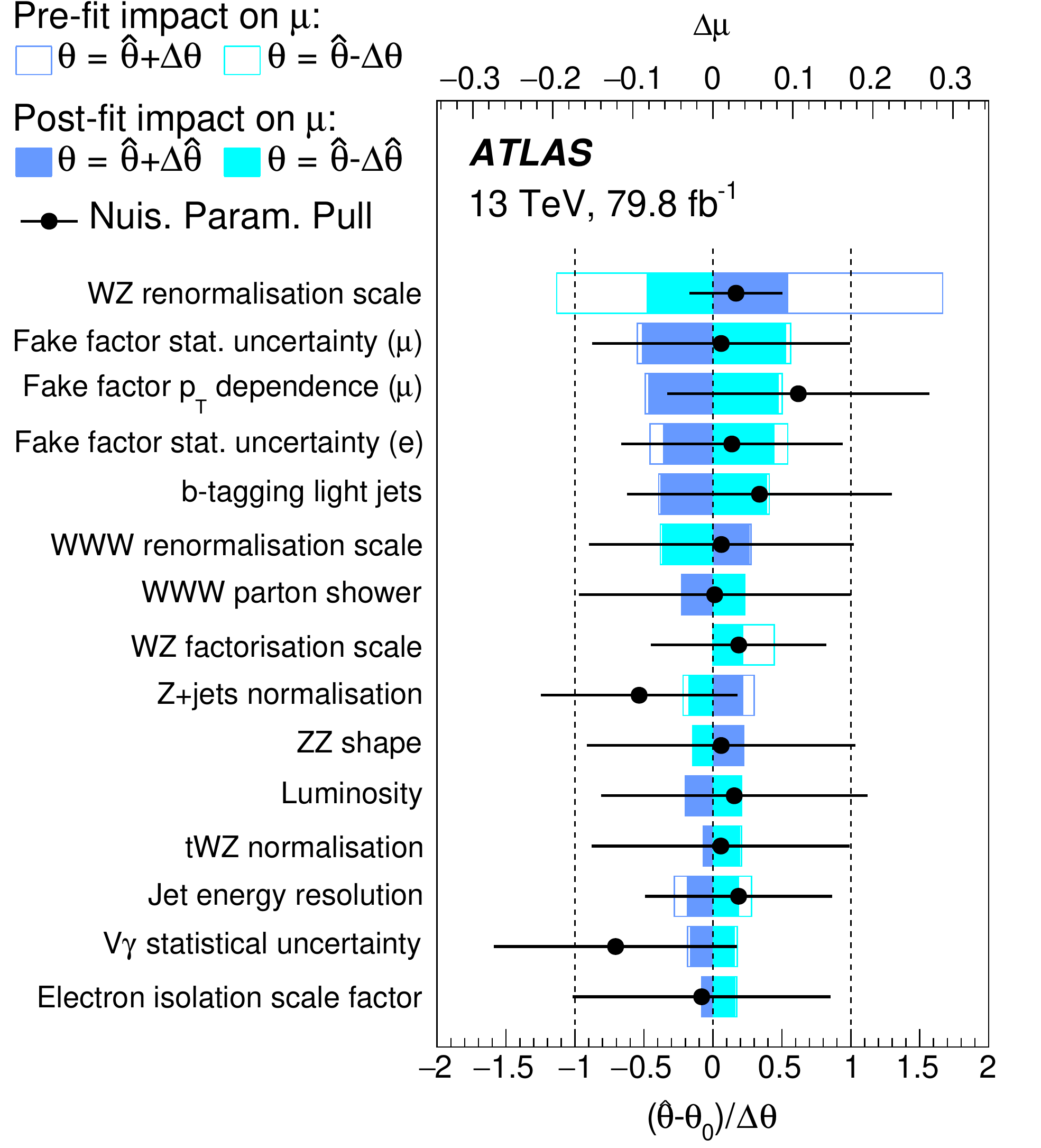}
\caption{
Impact of systematic uncertainties on the fitted signal-strength parameter $\mu$
for the combined $WVV$ fit to data. The systematic uncertainties are
listed in decreasing order of their post-fit impact in the fit, and only the
15 most important are displayed.
The effect of varying each nuisance parameter $\theta$ is shown, 
where $\theta_0$ is the pre-fit value, $\hat{\theta}$ is the post-fit value,
and $\Delta \theta$ and $\Delta \hat{\theta}$ are the pre- and post-fit uncertainties, 
respectively.
}
\label{fig:ranking}
\end{figure}

The overall observed (expected) significance for $WVV$
production is found to be 4.1$\sigma$ (3.1$\sigma$), constituting evidence for the
production of three massive vector bosons.
The combined best-fit signal strength for the $WVV$ process, obtained by the
fit to the eleven signal regions and one control region, is $\mu_{WVV} =
\obsvvverr$ with respect to the SM prediction (Section~\ref{s:detector}).  
The compatibility of the individual signal
strengths is \num{0.13}, determined by repeating the fit, assuming individual
signal strengths, and evaluating the $p$-value of the $\chi^2$ of the comparison.
The statistical uncertainty in the measured signal strength is $^{+0.25}_{-0.24}$ and the
systematic uncertainty is $^{+0.30}_{-0.27}$.  
The impact of the most important groups of systematic uncertainties on the measured value of
$\mu_{WVV}$ is shown in Table~\ref{tab:uncertainties}.
The largest systematic
uncertainties come from uncertainties related to data-driven background
evaluations affecting the \WWW channels, from
theoretical uncertainties related to renormalisation and factorisation scale
variations and experimental uncertainties.
The impact of each systematic uncertainty on the result is assessed
and the ranking for the nuisance parameters with the largest contribution 
to the uncertainty in $\mu_{WVV}$ is shown in Figure~\ref{fig:ranking}.

\begin{table}[htbp]
\caption{Summary of the effects of the most important groups of systematic
uncertainties in $\mu_{WVV}$: Uncertainties related to data-driven background
evaluations affecting the \WWW channels (data-driven); theoretical
uncertainties related to renormalisation and factorisation scale variations,
mostly in the diboson background, evaluated using simulations (theory);
experimental uncertainties in the signal and background evaluations
(instrumental); the statistical uncertainty in simulated events 
(MC stat.~uncertainty); and
modelling uncertainty evaluated by comparing different event generators
(generators).}
\label{tab:uncertainties}
\vspace{0.5cm}
\centering
\begin{tabular}{lccc}
\toprule \toprule
Uncertainty source         & \multicolumn{2}{c}{$\Delta\mu_{WVV}$} \\
\midrule
        Data-driven                                     & +0.14 & --0.14 \\
        Theory                                          & +0.15 & --0.13  \\
        Instrumental                                    & +0.12 & --0.09  \\
        MC stat.~uncertainty                            & +0.06 & --0.04 \\
        Generators                                      & +0.04 & --0.03 \\
\midrule
        Total systematic uncertainty           & +0.30 & --0.27 \\
\bottomrule \bottomrule
\end{tabular}
\end{table}

Additional fits are performed separately in the $WWW$ and the $WVZ$ channels. 
For these fits the other signal strength is fixed to its SM expectation.
For the fits of the \WWW channels,
the $WZ$ control region defined in Section~\ref{s:www} is used in the fit.
The inclusion of the $WZ$ control region helps constraining the overall
normalisation of the $WZ$+jets background, which in the combined fit is 
constrained by the \WVZ three-lepton signal regions.
The \ttZ control region is used in the \WVZ fit, however, it is not used in the \WWW fit. 
The observed (expected) significance is 3.2$\sigma$ (2.4$\sigma$) for $WWW$ production and 3.2$\sigma$ (2.0$\sigma$) for
$WVZ$ production.  

\begin{figure}[htbp]
\centering
\subfloat[]{\includegraphics[width=0.53\textwidth]{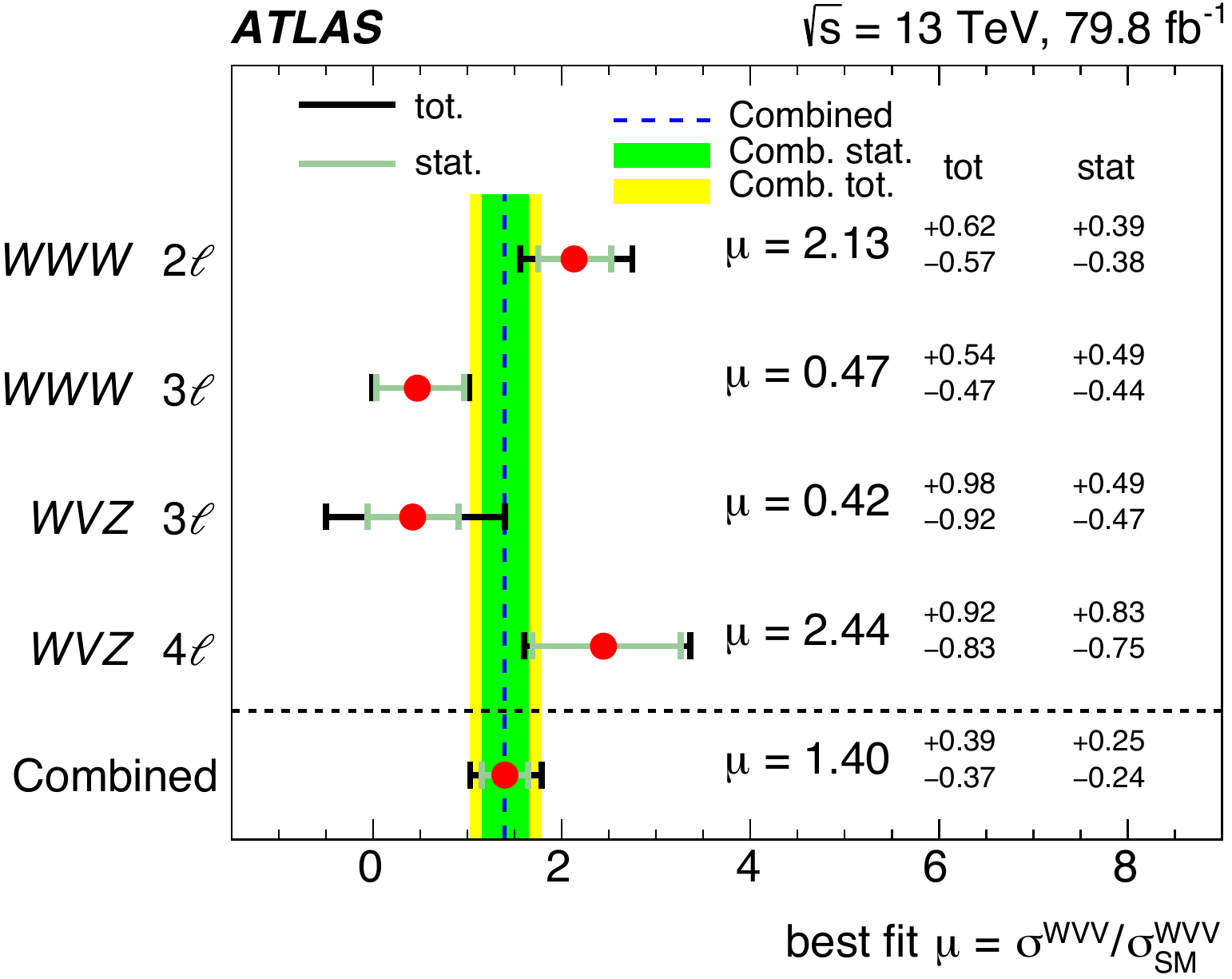}}
\hfill
\subfloat[]{\includegraphics[width=.44\textwidth]{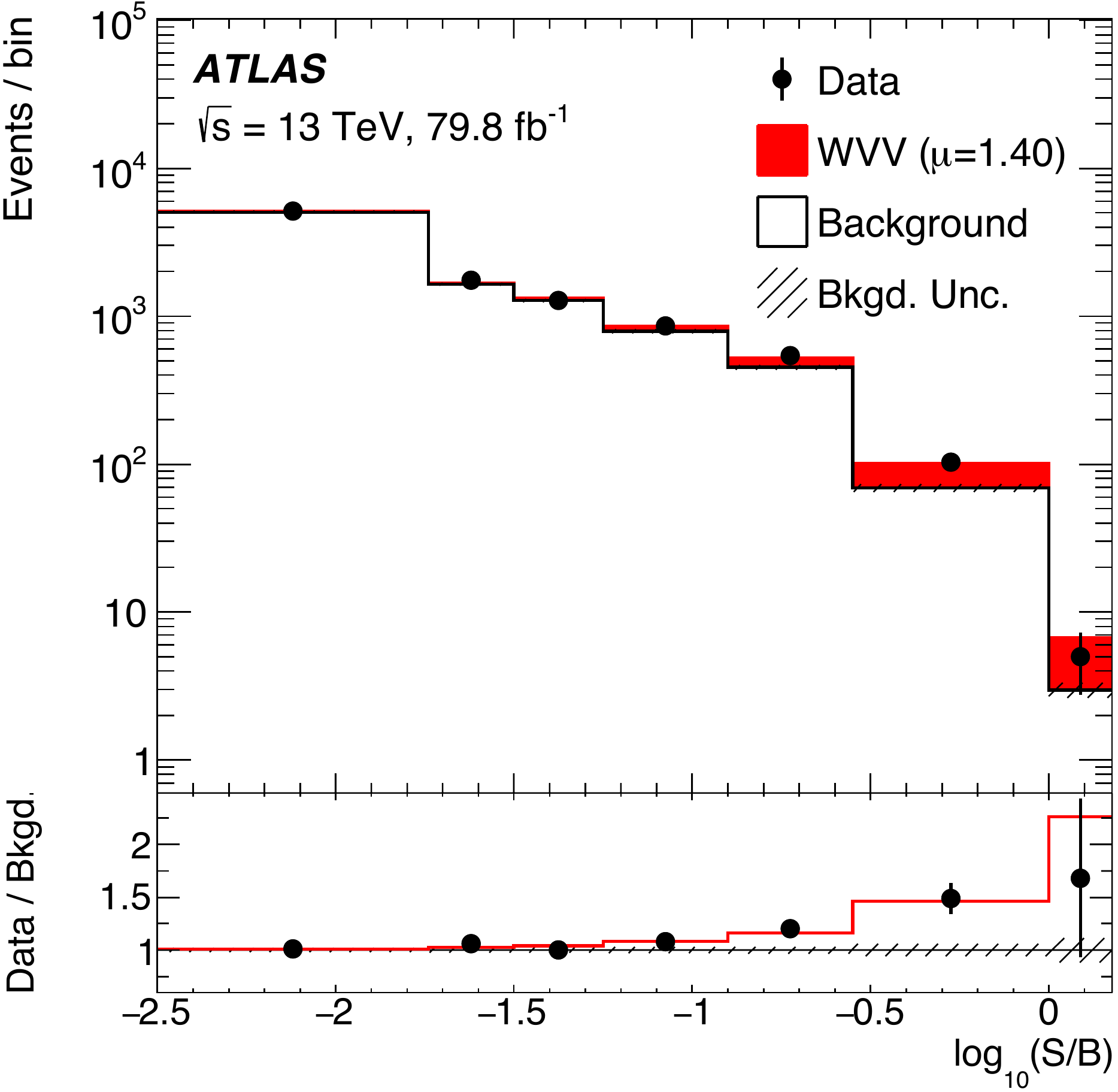}}
\caption{(a) Extracted signal strengths $\mu$ for the four analysis regions and
for the combination.  (b) Event yields as a function of
$\log_{10}\left(\text{S/B}\right)$ for data, background B and the signal S.
Events in all eleven signal regions are included. The background and signal
yields are shown after the global signal-plus-background fit.  The hatched band
corresponds to the systematic uncertainties, and the statistical uncertainties
are represented by the error bars on the data points.  The lower panel shows
the ratio of the data to the expected background estimated from the fit,
compared to the expected distribution including the signal (red line).
\label{fig:fig5}}
\end{figure}

Table~\ref{tab:sensitivity} and Figure~\ref{fig:fig5}(a) summarise the observed
and expected significances with respect to the background-only hypothesis and
the observed best-fit values of the signal strength for the individual and
combined fits. The measured signal strengths from the individual fits are
converted to inclusive cross-section measurements using the signal samples
described in Section~\ref{s:detector} and the central values of the theoretical
predictions.  All uncertainties determined in the fit are included in the
conversion, except for the normalisation uncertainty in the signal prediction.
The results are:
$\sigma_{WWW} = 0.65^{+0.16}_{-0.15}\,\text{(stat.)}\,^{+0.16}_{-0.14}\,\text{(syst.)}$ pb
and
$\sigma_{WWZ} = 0.55 \pm 0.14\,\text{(stat.)}\,^{+0.15}_{-0.13}\,\text{(syst.)}$ pb.  
For the
$\sigma_{WWZ}$ extraction, the $WZZ$ normalisation is fixed to the SM
expectation. The cross section of the latter is not reported, since there is
not enough sensitivity to this channel to quote a separate cross-section value.

\begin{table}
\caption{Observed and expected significances with respect to the SM
background-only hypothesis for the four $WVV$ channels entering the fit. 
\label{tab:sensitivity}}
\begin{center}
\begin{tabular}{l|cc}
\toprule 
\toprule 
\multirow{2}{*}{Decay channel} & \multicolumn{2}{c}{Significance}\\
 &     Observed       & Expected \\
\midrule
$WWW$ combined          & 3.2$\sigma$ & 2.4$\sigma$ \\
\qquad \sswww                  & 4.0$\sigma$ & 1.7$\sigma$ \\
\qquad \lllwww                 & 1.0$\sigma$ & 2.0$\sigma$ \\ 
\midrule
$WVZ$ combined          & 3.2$\sigma$ & 2.0$\sigma$ \\
\qquad \lllwvz                 & 0.5$\sigma$ & 1.0$\sigma$ \\
\qquad $\WVZ \to \ell \nu \ell \nu \ell \ell / qq \ell \ell \ell \ell$  & 3.5$\sigma$ & 1.8$\sigma$  \\ 
\midrule
$WVV$ combined                    & 4.1$\sigma$ & 3.1$\sigma$ \\
\bottomrule
\bottomrule
\end{tabular}
\end{center}
\end{table}

Figure~\ref{fig:fig5}(b) shows the data, background and signal yields, where
the discriminant bins in all signal regions are combined into bins of
$\log_{10}(\text{S/B})$, S being the expected signal yield and B the background
yield. The background and signal yields are shown after the global
signal-plus-background fit to the data.

\FloatBarrier

\section{Conclusion} 

In conclusion, a search for the joint production of three massive vector bosons
($W$ or $Z$) in proton--proton collisions using 79.8 fb$^{-1}$ of data at
$\sqrt{s} = \SI{13}{\TeV}$ collected by the ATLAS detector at the LHC, is
presented.  Events with two, three or four reconstructed electrons or muons
are analysed.  Evidence for the production of three massive vector bosons is
observed with a combined significance of 4.1 standard deviations, where the
expectation is 3.1 standard deviations.  The measured production cross sections
are 
$\sigma_{WWW} = 0.65^{+0.23}_{-0.21}$ pb, 
and 
$\sigma_{WWZ} = 0.55^{+0.21}_{-0.19}$ pb, 
in agreement with the Standard Model predictions.

\section*{Acknowledgements}
We thank CERN for the very successful operation of the LHC, as well as the
support staff from our institutions without whom ATLAS could not be operated
efficiently.

We acknowledge the support of ANPCyT, Argentina; YerPhI, Armenia; ARC,
Australia; BMWFW and FWF, Austria; ANAS, Azerbaijan; SSTC, Belarus; CNPq and
FAPESP, Brazil; NSERC, NRC and CFI, Canada; CERN; CONICYT, Chile; CAS, MOST and
NSFC, China; COLCIENCIAS, Colombia; MSMT CR, MPO CR and VSC CR, Czech Republic;
DNRF and DNSRC, Denmark; IN2P3-CNRS, CEA-DRF/IRFU, France; SRNSFG, Georgia;
BMBF, HGF, and MPG, Germany; GSRT, Greece; RGC, Hong Kong SAR, China; ISF and
Benoziyo Center, Israel; INFN, Italy; MEXT and JSPS, Japan; CNRST, Morocco;
NWO, Netherlands; RCN, Norway; MNiSW and NCN, Poland; FCT, Portugal; MNE/IFA,
Romania; MES of Russia and NRC KI, Russian Federation; JINR; MESTD, Serbia;
MSSR, Slovakia; ARRS and MIZ\v{S}, Slovenia; DST/NRF, South Africa; MINECO,
Spain; SRC and Wallenberg Foundation, Sweden; SERI, SNSF and Cantons of Bern
and Geneva, Switzerland; MOST, Taiwan; TAEK, Turkey; STFC, United Kingdom; DOE
and NSF, United States of America. In addition, individual groups and members
have received support from BCKDF, CANARIE, CRC and Compute Canada, Canada;
COST, ERC, ERDF, Horizon 2020, and Marie Sk{\l}odowska-Curie Actions, European
Union; Investissements d' Avenir Labex and Idex, ANR, France; DFG and AvH
Foundation, Germany; Herakleitos, Thales and Aristeia programmes co-financed by
EU-ESF and the Greek NSRF, Greece; BSF-NSF and GIF, Israel; CERCA Programme
Generalitat de Catalunya, Spain; The Royal Society and Leverhulme Trust, United
Kingdom. 

The crucial computing support from all WLCG partners is acknowledged
gratefully, in particular from CERN, the ATLAS Tier-1 facilities at TRIUMF
(Canada), NDGF (Denmark, Norway, Sweden), CC-IN2P3 (France), KIT/GridKA
(Germany), INFN-CNAF (Italy), NL-T1 (Netherlands), PIC (Spain), ASGC (Taiwan),
RAL (UK) and BNL (USA), the Tier-2 facilities worldwide and large non-WLCG
resource providers. Major contributors of computing resources are listed in
Ref.~\cite{ATL-GEN-PUB-2016-002}.

\clearpage

\printbibliography
\clearpage

 
\begin{flushleft}
{\Large The ATLAS Collaboration}

\bigskip

G.~Aad$^\textrm{\scriptsize 101}$,    
B.~Abbott$^\textrm{\scriptsize 128}$,    
D.C.~Abbott$^\textrm{\scriptsize 102}$,    
O.~Abdinov$^\textrm{\scriptsize 13,*}$,    
A.~Abed~Abud$^\textrm{\scriptsize 70a,70b}$,    
K.~Abeling$^\textrm{\scriptsize 53}$,    
D.K.~Abhayasinghe$^\textrm{\scriptsize 93}$,    
S.H.~Abidi$^\textrm{\scriptsize 167}$,    
O.S.~AbouZeid$^\textrm{\scriptsize 40}$,    
N.L.~Abraham$^\textrm{\scriptsize 156}$
% ...
\bigskip
\\

% ...
$^{13}$Institute of Physics, Azerbaijan Academy of Sciences, Baku; Azerbaijan.\\
% ...
$^{40}$Niels Bohr Institute, University of Copenhagen, Copenhagen; Denmark.\\
% ...
$^{53}$II. Physikalisches Institut, Georg-August-Universit\"{a}t G\"ottingen, G\"ottingen; Germany.\\
% ...
$^{70}$$^{(a)}$INFN Sezione di Pavia;$^{(b)}$Dipartimento di Fisica, Universit\`a di Pavia, Pavia; Italy.\\
% ...
$^{93}$Department of Physics, Royal Holloway University of London, Egham; United Kingdom.\\
% ...
$^{101}$CPPM, Aix-Marseille Universit\'e, CNRS/IN2P3, Marseille; France.\\
$^{102}$Department of Physics, University of Massachusetts, Amherst MA; United States of America.\\
% ...
$^{128}$Homer L. Dodge Department of Physics and Astronomy, University of Oklahoma, Norman OK; United States of America.\\
% ...
$^{156}$Department of Physics and Astronomy, University of Sussex, Brighton; United Kingdom.\\
% ...
$^{167}$Department of Physics, University of Toronto, Toronto ON; Canada.\\
% ...
$^{a}$ Also at Borough of Manhattan Community College, City University of New York, New York NY; United States of America.\\
$^{b}$ Also at Centre for High Performance Computing, CSIR Campus, Rosebank, Cape Town; South Africa.\\
% ...
$^{*}$ Deceased

\end{flushleft}


\end{document}